\documentclass[12pt]{article}
\usepackage{ctex}
\usepackage{amssymb,amsmath,color,amsmath,bm}
\usepackage[colorlinks,linkcolor=blue]{hyperref}
\usepackage{geometry}
\usepackage{tabularx}
\usepackage{bm}
\usepackage{authblk}
\usepackage[square, comma, sort&compress, numbers]{natbib}
\usepackage{multirow}
\usepackage{float}  

	\newcommand{\rank}{{\mathrm{rank}}}
	\newtheorem{theorem}{Theorem}[section]
	\newtheorem{definition}{Definition}[section]
	\newtheorem{lemma}{Lemma}[section]
	\newtheorem{example}{Example}[section]
	
	\newtheorem{corollary}{Corollary}[section]
	\newtheorem{remark}{Remark}[section]

	\catcode`@=11 \@addtoreset{equation}{section} \catcode`@=12
\begin{document}
\title{\bf Two classes of NMDS codes from Roth-Lempel codes\thanks{This paper is supported by National Natural Science Foundation of China (12471494) and Natural Science Foundation of Sichuan Province (2024NSFSC2051). The corresponding author is professor Qunying Liao.}}
\author{\small Zhonghao Liang}
\author{\small Qunying Liao
	{\thanks{E-mail:liangzhongh0807@163.com; qunyingliao@sicnu.edu.cn}}
}
\affil[] {\small(School of Mathematical Sciences, Sichuan Normal University, Chengdu, 610066, China)}
\date{}
\maketitle
	{\bf Abstract.}
	{\small Since near maximum distance separable (NMDS) codes have good algebraic properties and excellent error-correcting capabilities, they have been widely used in various fields such as communication systems, data storage, quantum codes, and so on. In this paper, basing on the generator matrix of  Roth-Lempel codes, we present
		two classes of NMDS codes which generalize Han's and Zheng's constructions in 2023 and 2025, respectively. And we also completely determine their weight
		distributions.
 }\\
	

\section{Introduction}
Let $\mathbb{F}_q$ be the finite field with $q=p^m$ elements and $\mathbb{F}_{q}^{*}=\mathbb{F}_{q}\backslash\left\{0\right\}$, where $p$ is prime. An $[n,k,d]$ linear code $\mathcal{C}$ over $\mathbb{F}_q$ is a $k$-dimensional subspace of $\mathbb{F}_q^n$ with minimum (Hamming) distance $d$ and length $n$. The dual code of $\mathcal{C}$
is defined as
\[
\mathcal{C}^{\perp}=\left\{\left(x_{1}, \ldots, x_{n}\right)=\boldsymbol{x}\in\mathbb{F}_q^{n} \mid\langle\boldsymbol{x},\boldsymbol{y}\rangle=\sum\limits_{i=1}^{n} x_{i} y_{i}=0,  \forall \boldsymbol{y}=\left(y_{1}, \ldots, y_{n}\right) \in \mathcal{C}\right\}.
\]
Let $A_i(i=1,\ldots,n)$ be the number of codewords with Hamming weight $i$ in $\mathcal{C}$, the weight enumerator of $\mathcal{C}$ is defined by the polynomial $1+A_1z+\cdots+A_nz^n$ and the sequence $(1,A_1,\ldots,A_n)$ is called its
weight distribution. The weight distribution is an important parameter for a linear code, which can
be applied to determine the capability for both error-detection and error-correction\cite{A1}.

For an $[n, k, d]_q$ linear code $\mathcal{C}$, the Singleton bound of $\mathcal{C}$ is
defined by the non-negative integer  $S(\mathcal{C})=n−k+1−d$. If $S(\mathcal{C}) =0$, then $\mathcal{C}$ is maximum distance separable (in short, MDS). If $S(\mathcal{C})=1$, then $\mathcal{C}$ is almost MDS (in short, AMDS). If $S(\mathcal{C}) = S(\mathcal{C}^{\perp})=1$, then $\mathcal{C}$ is near MDS (in short, NMDS).

The Reed-Solomon code, as a good class  of MDS linear codes, is defined as
$$
\mathrm{RS}_{k}(\boldsymbol{\alpha}):=\left\{\left(f\left(\alpha_{1}\right), \ldots,f\left(\alpha_{n}\right)\right) | f(x) \in \mathbb{F}_{q}^{k}[x]\right\},
$$
where $\mathbb{F}_{q}[x]$ is the polynomial ring over $\mathbb{F}_{q}$,
$$\mathbb{F}_{q}^{k}[x]=\left\{f(x)=\sum_{i=0}^{k-1} f_{i} x^{i}\mid f_{i} \in \mathbb{F}_{q}, 0 \leq i \leq k-1\right\},$$
and $\boldsymbol{\alpha}=\left(\alpha_{1}, \ldots, \alpha_{n}\right) \in \mathbb{F}_{q}^{n}$ with $\alpha_{i} \neq \alpha_{j}(i \neq j)$.
It's easy to prove that  
\begin{equation}\label{RS}
	\boldsymbol{G}_{RS}=\left(\begin{matrix}
		 1&		1&		\cdots&		1&1\\
		 \alpha _1&\alpha _2&		\cdots&\alpha _{n-1}&\alpha _n\\
		 \vdots&		\vdots&		\quad&		\vdots&		\vdots\\
		 \alpha _{1}^{k-2}&\alpha _{2}^{k-2}&		\cdots&\alpha _{n-1}^{k-2}&\alpha _{n}^{k-2}\\
		 \alpha _{1}^{k-1}&\alpha _{2}^{k-1}&		\cdots&\alpha _{n-1}^{k-1}&\alpha _{n}^{k-1}\\
	\end{matrix} \right)_{k\times n}
\end{equation}	
is a generator matrix of $\mathrm{RS}_{k}(\boldsymbol{\alpha})$ and $\mathrm{RS}_{k}(\boldsymbol{\alpha})$ has the parameters $\left[n,k,n-k+1\right]_{q}$. In addition, if a linear code $\mathcal{C}$ is not equivalent to any RS code, then  $\mathcal{C}$ is the non-Reed-
Solomon (non-RS) type and  $\mathcal{C}$  is called to be a non-RS code. 

Since both MDS and NMDS codes are very important in coding theory and applications \cite{A2,A3,A4,A5,A6}, and so it's interesting to construct MDS or NMDS codes and determine their weight distributions \cite{A7,A8,A9,A10,A11,A12,A13,A14,A15,A16,A17,A23}.
In 1989, Roth and Lempel \cite{A18} constructed a class of non-RS MDS codes by adding two columns to  the matrix $\boldsymbol{G}_{RS}$ given by $(\ref{RS})$, the corresponding linear code over $\mathbb{F}_{q}$ has the generator matrix 
\begin{equation}\label{G_2}
\left(\begin{matrix}
		\boldsymbol{G}_{RS}&\begin{matrix}
			\boldsymbol{0}_{(k-2)\times 2}\\
			\boldsymbol{M}_{\delta}
		\end{matrix}
	\end{matrix} \right)_{k\times (n+2)},
\end{equation} 
where $4\leq k+1\leq n\leq q$, $\boldsymbol{M}_{\delta}=\left(\begin{matrix}
	0&1\\
	1&\delta
\end{matrix}\right)$ with $\delta\in\mathbb{F}_{q}$, and denote by $\mathrm{RL}_{k}(\boldsymbol{\alpha},\boldsymbol{M}_{\delta})$. Especially, when $\delta=0$, Han and Fan \cite{A16} presented a necessary and sufficient condition for the code $\mathrm{RL}_{k}(\boldsymbol{\alpha},\boldsymbol{M}_{0})$ to be NMDS, proved that $\mathrm{RL}_{k}(\mathbb{F}_{q},\boldsymbol{M}_{0})$ is NMDS for the case $3\leq k\leq q$, and determined its weight distribution. In 2024, Ding and Zhu \cite{A12} constructed two classes of NMDS codes and  determined their weight distributions, the corresponding linear codes over $\mathbb{F}_{2^m}$ have the generator matrix
$$\begin{pmatrix}
	\begin{matrix}
	1 & 1 & \cdots & 1 \\
	\alpha_{1} & \alpha_{2} & \cdots & \alpha_{2^m-1} \\
	\alpha_{1}^{2^{h}} & \alpha_{2}^{2^{h}} & \cdots & \alpha_{2^m-1}^{2^{h}}\\
	\alpha_{1}^{2^{h}+1} & \alpha_{2}^{2^{h}+1} & \cdots & \alpha_{2^m-1}^{2^{h}+1}\\
	\end{matrix}&\begin{matrix}
	\boldsymbol{0}_{2\times 2}\\
	\\ 
	\boldsymbol{M}_{0}
	\end{matrix}
	
	\end{pmatrix}_{4\times (2^m+1)}$$
and	
$$\begin{pmatrix}
\begin{matrix}
1 & 1 & \cdots & 1\\
\alpha_{1} & \alpha_{2} & \cdots & \alpha_{2^m} \\
\alpha_{1}^{2^{h}} & \alpha_{2}^{2^{h}} & \cdots & \alpha_{2^m}^{2^{h}}\\
\alpha_{1}^{2^{h}+1} & \alpha_{2}^{2^{h}+1} & \cdots & \alpha_{2^m}^{2^{h}+1}\\
\end{matrix}&\begin{matrix}
	\boldsymbol{0}_{2\times 2}\\
	\\ 
	\boldsymbol{M}_{0}
\end{matrix}
\end{pmatrix}_{4\times (2^m+2)},$$
respectively. Especially, when $h=1$, the corresponding linear codes are $\mathrm{RL}_{4}\left(\mathbb{F}_{2^m}^{*},\boldsymbol{M}_{0}\right)$ and $\mathrm{RL}_{4}\left(\mathbb{F}_{2^m},\boldsymbol{M}_{0}\right)$, respectively. Recently, Zhang and Zheng \cite{A13} generalized Zhu's construction, i.e., they proved that $\mathrm{RL}_{k}\left(\mathbb{F}_{q}^{*},\boldsymbol{M}_{0}\right)$ is NMDS for $4\leq k\leq q-3$ and $p=2$, or $3\leq k\leq 4$ and $p\neq 2$, the corresponding linear code over $\mathbb{F}_{q}$ has the generator matrix 
$$\begin{pmatrix}
\begin{matrix}
	1 & \cdots & 1\\
	\alpha_{1} & \cdots & \alpha_{q-1} \\
	\vdots & \ddots & \vdots\\
	\alpha_{1}^{k-4} & \cdots & \alpha_{q-1}^{k-4}\\
	\alpha_{1}^{k-3} & \cdots & \alpha_{q-1}^{k-3}\\
\end{matrix}&\boldsymbol{0}_{(k-2)\times 2}\\
\begin{matrix}
\alpha_{1}^{k-1} & \cdots & \alpha_{q-1}^{k-1}\\
\alpha_{1}^{k-2} & \ldots & \alpha_{q-1}^{k-2}
\end{matrix}&\boldsymbol{M}_{0}\\
\end{pmatrix}_{k\times (q+1)}.$$

We continue the above work, i.e., replace $\boldsymbol{M}_{0}$ by $\boldsymbol{M}_{2\times 2}=(a_{ij})\in\mathrm{GL}_{2}\left(\mathbb{F}_{q}\right)$ and generalize the corresponding results.

This paper is organized as follows. In Section 2, we give the definition of the generalized Roth-Lempel code and some necessary lemmas. In Sections 3-4, we prove that  $\mathrm{RL}_{k}\left(\mathbb{F}_{q}^{*},\boldsymbol{M}_{2\times 2}\right)$ and $\mathrm{RL}_{k}\left(\mathbb{F}_{q},\boldsymbol{M}_{2\times 2}\right)$ are both NMDS and then  completely determine their weight distributions. In section 5, we conclude the whole paper.
	\section{Preliminaries}
In this section, we give the definition of the generalized Roth-Lempel code and recall some necessary lemmas.
\begin{definition}\label{EGRLdefinition}
	Let $\mathbb{F}_q$ be the finite field of $q$ elements, where $q$ is a prime power. Let $l+1\leq k+1\leq n\leq q$,  $\boldsymbol{\alpha}=\left(\alpha_{1}, \ldots, \alpha_{n}\right) \in \mathbb{F}_{q}^{n}$ with $\alpha_{i} \neq \alpha_{j}(i \neq j)$, $b\in\mathbb{F}_{q}^{*}$ and $\boldsymbol{v}=$ $\left(v_{1}, \ldots, v_{n}\right) \in\left(\mathbb{F}_{q}^{*}\right)^{n}$. 
	The generalized Roth-Lempel (in short, GRL) code $\mathrm{GRL}_{k}(\boldsymbol{\alpha}, \boldsymbol{v},\boldsymbol{M}_{l\times l})$ is defined as
	$$
	\mathrm{GRL}_{k}(\boldsymbol{\alpha}, \boldsymbol{v},\boldsymbol{M}_{l\times l})\triangleq\left\{\left(v_{1} f\left(\alpha_{1}\right), \ldots, v_{n} f\left(\alpha_{n}\right),\boldsymbol{\beta}\right) | f(x) \in \mathbb{F}_{q}^{k}[x]\right\},
	$$
	where $\boldsymbol{M}_{l\times l}=(a_{ij})_{l\times l}\in\mathrm{GL}_{l}\left(\mathbb{F}_{q}\right)$ and  $$\begin{aligned}
	\boldsymbol{\beta}=&\left(f_{k-l},\ldots,f_{k-1}\right)\boldsymbol{M}_{l\times l}\\
	=&\left(a_{11}f_{k-l}+a_{21}f_{k-(l-1)}+\cdots+a_{l1}f_{k-1},\ldots,a_{1l}f_{k-l}+a_{2l}f_{k-(l-1)}+\cdots+a_{ll}f_{k-1}\right).
	\end{aligned}$$
\end{definition} 
\begin{remark}\label{generatormatrix}
By taking $\boldsymbol{v}=(1,\ldots,1)\in\mathbb{F}_{q}^{n}$ and $\boldsymbol{M}_{l\times l}=\begin{pmatrix} 
0&1\\
1&\delta
\end{pmatrix}$ or $\begin{pmatrix} 
0&0&1\\
0&1&\tau\\
1&\delta&\pi\\
\end{pmatrix}$ in Definition $\ref*{EGRLdefinition}$, the corresponding code is just the classical Roth-Lempel code or $\overline{\mathrm{RL}_{k}\left(\boldsymbol{\alpha},\delta,k,n+2\right)}(\boldsymbol{u})$ in \cite{A18} or \cite{A22}, respectively.

%
\end{remark}

\begin{lemma}\label{MDSdefinition}
	{\rm(\cite{A3}, Lemma 3.7)}
Let $\mathcal{C}$ be an $[n, k]$ code over $\mathbb{F}_{q}$ with $k\geq 1$.  $\boldsymbol{G}$  and  $\boldsymbol{H}$  are the 
generator matrix and parity-check matrix for $\mathcal{C}$, respectively. Then, the following statements are equivalent to each other,

$(1)\ \mathcal{C}$ is MDS;

$(2)$ any $k$ columns of $\boldsymbol{G}$ are linearly independent over $\mathbb{F}_{q}$;

$(3)$ any $n-k$ columns of  $\boldsymbol{H}$  are linearly independent over $\mathbb{F}_{q}$;

$(4)\  \mathcal{C}^{\perp}$ is MDS.
\end{lemma}

The following lemma provides the crucial information on the weight distributions of NMDS codes and their duals.
\begin{lemma}\label{NMDSweight}{\rm(\cite{A16}, Lemma 2)} Let $\mathcal{C}$ be an $[n, k, n-k]$ NMDS code over $\mathbb{F}_q$. Then $\mathcal{C}$ and $\mathcal{C}^{\perp}$ have the same number of
	minimum weight codewords, i.e., $A_{n-k}=A_{k}^{\perp}.$ Moreover, the
	weight distributions of $\mathcal{C}$ and $\mathcal{C}^{\perp}$ are given by
	\begin{align}\label{ANMDS}
	A_{n-k+\ell}=\binom{n}{k-\ell}\sum\limits_{j=0}^{\ell-1}(-1)^j\binom{n-k+\ell}{j}(q^{\ell-j}-1)\!+\!(-1)^{\ell}\binom{k}{\ell}A_{n-k}(1\leq\ell\leq k),
	\end{align}
and \begin{align}\label{DNMDS}
	A_{k+\ell}^{\perp}=\binom{n}{k+\ell}\sum\limits_{j=0}^{\ell-1}(-1)^j\binom{k+\ell}{j}(q^{\ell-j}-1)\!+\!(-1)^{\ell}\binom{n-k}{\ell}A_{k}^{\perp}(1\leq\ell\leq n-k),
\end{align}
respectively.
\end{lemma}

It's well-known that the subset sum problem is an $\mathbf{NP}$-complete problem. Given $k\in\mathbb{Z}^{+}$, $b\in\mathbb{F}_q$ and $D\subseteq\mathbb{F}_q$ with $|D|\ge k$, to determine the value of 
\begin{align*}
\# N(k,b,D)=\# \left\{\{x_1,\ldots,x_k\}\subseteq D\big|x_1+\cdots+x_k=b\right\}
\end{align*}
is the subset sum problem. 

For $D=\mathbb{F}_q$ or $\mathbb{F}_q^{*}$, $\#N(t,b,D)$ is given explicitly by the following 
\begin{lemma}\label{subsetsum}
{\rm(\cite{A20}, Theorem 1.2)} For any $b\in\mathbb{F}_{q}$, define $v(b)=\begin{cases}
	-1,&\text{if}\ b\neq 0;\\
	q-1,&\text{if}\ b=0,\\
\end{cases}$ then		
	\begin{align*}
	\#N(k,b,\mathbb{F}_{q}^{*})=
	\frac{1}{q}\binom{q-1}{k}+(-1)^{k+\lfloor\frac{k}{p}\rfloor}\frac{v(b)}{q}\binom{\frac{q}{p}-1}{\lfloor\frac{k}{p}\rfloor},
	\end{align*}
and 
$$\# N(k,b,\mathbb{F}_{q})=\begin{cases}
	\frac{1}{q}\binom{q}{k},&\text{if}\ p\nmid k;\\
	\frac{1}{q}\binom{q}{k}+(-1)^{k+\frac{k}{p}}\frac{v(b)}{q}\binom{\frac{q}{p}}{\frac{k}{p}}, &\text{if}\ p\mid k.\\
\end{cases}$$
\end{lemma}
\begin{lemma}\label{subsetsumneq0}
{\rm(\cite{A21}, Remark 2.6)} If $k \geq 2$ and $D=\mathbb{F}_{q}^{*}$ or $\mathbb{F}_{q}$, then $\# N\left(k, b,D\right)=0$ if and only if $2 | q$ and $(k, b) \in\{(2,0),(q-2,0)\}$.
\end{lemma}

By the proof of Lemma 6 in \cite{A13}, we have the following Lemma \ref{subsetsum0}.
\begin{lemma}\label{subsetsum0}
{\rm(\cite{A13})} $\# N(k-1, 0, \mathbb{F}_q^*) = 0$ if and only if $k \in \{3, q-2, q-1\}$ and $p=2$, or $k \in \{2, q-1\}$ and $p \neq 2$.
\end{lemma} 

\begin{lemma}\label{Vandermonde}
	{\rm(\cite{A17}, Lemma 6)} Let $x_1,\ldots,x_n$ be pairwise distinct elements of $\mathbb{F}_q$, then
	\begin{align*}
		\det\left(\begin{array}{ccccc}
			1 &  \ldots &  1 \\ 
			x_1  & \ldots &  x_n \\
			\vdots &  \ddots &  \vdots \\
			x_1^{n-2} &  \ldots &  x_n^{n-2} \\
			x_1^n &  \ldots &  x_n^n 
		\end{array}\right)=\sum_{t=1}^{n}x_t\prod_{1\leq i<j\leq n}(x_j-x_i).             
	\end{align*}
\end{lemma}

\section{The NMDS property of $\text{RL}_{k}\left(\mathbb{F}_{q}^{*},M_{2\times 2}\right)$}\label{sec3}
Let $\mathbb{F}_{q}^{*}=\left\{\beta_1,\ldots,\beta_{q-1}\right\}$, in this section, we prove that  $\mathrm{RL}_{k}\left(\mathbb{F}_{q}^{*},\boldsymbol{M}_{2\times 2}\right)$ is NMDS,
and determine its weight distribution for $4\leq k\leq q-3$ and $p=2$,  or $3\leq k\leq q-2$ and $ p\neq 2$. Since their proofs are a little long, for the convenience, we first determine the parameters of $\mathrm{RL}_{k}^{\perp}\left(\mathbb{F}_{q}^{*},\boldsymbol{M}_{2\times 2}\right)$.

\begin{lemma}\label{C1dualdistance}
For any positive integer $k\geq 3$, we have $d\left(\mathrm{RL}_{k}^{\perp}\left(\mathbb{F}_{q}^{*},\boldsymbol{M}_{2\times 2}\right)\right) \geq k$.
\end{lemma}
{\bf Proof}. By Definition \ref{EGRLdefinition}, it's easy to show that $\mathrm{RL}_{k}^{\perp}\left(\mathbb{F}_{q}^{*},\boldsymbol{M}_{2\times 2}\right)$ has the parity-check matrix
\begin{equation}\label{FirstNMDS}
	\boldsymbol{G}_{1}=\begin{pmatrix}
		1 & \cdots & 1 &0& 0\\
		\beta_{1} & \cdots & \beta_{q-1} & 0 & 0 \\
		\vdots & \ddots & \vdots & \vdots & \vdots \\
		\beta_{1}^{k-4} & \cdots & \beta_{q-1}^{k-4} & 0 & 0 \\
		\beta_{1}^{k-3} & \cdots & \beta_{q-1}^{k-3} & 0 & 0 \\
		\beta_{1}^{k-2} & \ldots & \beta_{q-1}^{k-2} &  a_{11} &  a_{12} \\
		\beta_{1}^{k-1} & \cdots & \beta_{q-1}^{k-1} & a_{21} & a_{22} \\
	\end{pmatrix}_{k\times (q+1)},
\end{equation}
where $\boldsymbol{M}_{2\times 2}=\begin{pmatrix}
	a_{11} & a_{12} \\
	a_{21} &  a_{22} \\
\end{pmatrix}\in\mathrm{GL}_{2}(\mathbb{F}_{q}).$

For the convenience, we set  $$\boldsymbol{u}_{1}=\left(0,\ldots,0,a_{11},a_{21}\right)^{T},\boldsymbol{u}_{2}=\left(0,\ldots,0,a_{12},a_{22}\right)^{T}.$$ 
By Corollary 1.4.14 of \cite{A3}, it's enough to prove that any $k-1$ columns of $\boldsymbol{G}_{1}$ are $\mathbb{F}_{q}$-linearly
	independent, i.e., the submatrix consisted of any $k-1$ columns in $\boldsymbol{G}_{1}$ is nonsingular over $\mathbb{F}_{q}$. Thus we have the following three cases.
	
	\textbf{Case 1}.  Assume that the submatrix $\boldsymbol{B}_{1}$ consisted of $k-1$ columns in $\boldsymbol{G}_{1}$ contains neither $\boldsymbol{u}_{1}$ nor $\boldsymbol{u}_{2}$ , i.e.,
	$$\boldsymbol{B}_{1}=\begin{pmatrix} 
		1 & \cdots & 1\\
		\beta_{i_1} & \cdots & \beta_{i_{k-1}} \\
		\vdots & \ddots & \vdots \\
		\beta_{i_1}^{k-4} & \cdots & \beta_{i_{k-1}}^{k-4} \\
		\beta_{i_1}^{k-3} & \cdots & \beta_{i_{k-1}}^{k-3} \\
		\beta_{i_1}^{k-2} & \ldots & \beta_{i_{k-1}}^{k-2} \\
		\beta_{i_1}^{k-1} & \cdots & \beta_{i_{k-1}}^{k-1} 
	\end{pmatrix}_{k\times(k-1)},$$
	we consider the matrix $\boldsymbol{D}_1$ given by deleting the last row of  $\boldsymbol{B}_{1}$, i.e.,
	$$\boldsymbol{D}_{1}=\begin{pmatrix} 
		1 & \cdots & 1\\
		\beta_{i_1} & \cdots & \beta_{i_{k-1}} \\
		\vdots & \ddots & \vdots \\
		\beta_{i_1}^{k-4} & \cdots & \beta_{i_{k-1}}^{k-4} \\
		\beta_{i_1}^{k-3} & \cdots & \beta_{i_{k-1}}^{k-3} \\
		\beta_{i_1}^{k-2} & \ldots & \beta_{i_{k-1}}^{k-2} 
	\end{pmatrix}_{(k-1)\times(k-1)},$$
	then 
	$$\det (\boldsymbol{D}_{1})=\prod\limits_{1 \leq j<l\leq k-1}\left(\beta _{i_l}-\beta _{i_j}\right).$$
	Note that $\prod\limits_{1 \leq j< l\leq k-1}\left(\beta_{i_{l}}-\beta_{i_{j}}\right)\neq 0,$
	hence, the $k-1$ columns of $\boldsymbol{B}_{1}$ are $\mathbb{F}_{q}$-linearly independent.
	
	\textbf{Case 2}. Assume that the submatrix $\boldsymbol{B}_{2}$ consisted of $k-1$ columns in $\boldsymbol{G}_{1}$ contains either    $\boldsymbol{u}_{1}$ or $\boldsymbol{u}_{2}$, i.e.,
	$$\boldsymbol{B}_{2}=\begin{pmatrix} 
		1 & \cdots & 1&0\\
		\beta_{i_1} & \cdots & \beta_{i_{k-2}}&0 \\
		\vdots & \ddots & \vdots \\
		\beta_{i_1}^{k-4} & \cdots & \beta_{i_{k-2}}^{k-4}&0 \\
		\beta_{i_1}^{k-3} & \cdots & \beta_{i_{k-2}}^{k-3}&0 \\
		\beta_{i_1}^{k-2} & \ldots & \beta_{i_{k-2}}^{k-2}&a_{1s} \\
		\beta_{i_1}^{k-1} & \cdots & \beta_{i_{k-2}}^{k-1}&a_{2s} \\
	\end{pmatrix}_{k\times(k-1)}(s=1,2),$$
	Note that $\boldsymbol{M}_{2\times 2}\in\mathrm{GL}_{2}\left(\mathbb{F}_{q}\right)$, it means that $a_{1s}$ and $a_{2s}$ can not be zero, simultaneously. Without loss of generality, we set $a_{1s}\neq 0$. Now, we consider the matrix $\boldsymbol{D}_2$ given by deleting the last row of  $\boldsymbol{B}_{2}$, i.e.,
	$$\boldsymbol{D}_{2}=\begin{pmatrix} 
		1 & \cdots & 1&0\\
		\beta_{i_1} & \cdots & \beta_{i_{k-2}}&0 \\
		\vdots & \ddots & \vdots \\
		\beta_{i_1}^{k-4} & \cdots & \beta_{i_{k-2}}^{k-4}&0 \\
		\beta_{i_1}^{k-3} & \cdots & \beta_{i_{k-2}}^{k-3}&0 \\
		\beta_{i_1}^{k-2} & \ldots & \beta_{i_{k-2}}^{k-2}&a_{1s} \\ 
	\end{pmatrix}_{(k-1)\times(k-1)}(s=1,2),$$
	then 
	$$\det (\boldsymbol{D}_{2})= a_{1s}\prod\limits_{1 \leq j<l\leq k-2}\left(\beta _{i_l}-\beta _{i_j}\right).$$
	Note that $\prod\limits_{1 \leq j< l\leq k-2}\left(\beta_{i_{l}}-\beta_{i_{j}}\right)\neq 0,$
	hence, the $k-1$ columns of $\boldsymbol{B}_{2}$ are $\mathbb{F}_{q}$-linearly independent.
	
	\textbf{Case 3}. Assume that the submatrix $\boldsymbol{B}_{3}$ consisted of $k-1$ columns in $\boldsymbol{G}_{1}$ contains both $\boldsymbol{u}_{1}$ and $\boldsymbol{u}_{2}$, i.e.,
	$$\boldsymbol{B}_{3}=\begin{pmatrix} 
		1 & \cdots & 1&0 &0\\
		\beta_{i_1} & \cdots & \beta_{i_{k-3}}&0&0 \\
		\vdots & \ddots & \vdots& \vdots& \vdots \\
		\beta_{i_1}^{k-4} & \cdots & \beta_{i_{k-3}}^{k-4}&0&0 \\
		\beta_{i_1}^{k-3} & \cdots & \beta_{i_{k-3}}^{k-3}&0&0 \\
		\beta_{i_1}^{k-3} & \ldots & \beta_{i_{k-3}}^{k-2}&a_{11}&a_{12} \\
		\beta_{i_1}^{k-1} & \cdots & \beta_{i_{k-3}}^{k-1}&a_{21}&a_{22} \\
	\end{pmatrix}_{k\times(k-1)},$$
	we consider the matrix $\boldsymbol{D}_3$ given by deleting the fist row of  $\boldsymbol{B}_{3}$, i.e.,
	$$\boldsymbol{D}_{3}=\begin{pmatrix} 
		\beta_{i_1} & \cdots & \beta_{i_{k-3}}&0&0 \\
		\vdots & \ddots & \vdots& \vdots& \vdots \\
		\beta_{i_1}^{k-4} & \cdots & \beta_{i_{k-3}}^{k-4}&0&0 \\
		\beta_{i_1}^{k-3} & \cdots & \beta_{i_{k-3}}^{k-3}&0&0 \\
		\beta_{i_1}^{k-3} & \ldots & \beta_{i_{k-3}}^{k-2}&a_{11}&a_{12} \\
		\beta_{i_1}^{k-1} & \cdots & \beta_{i_{k-3}}^{k-1}&a_{21}&a_{22} \\
	\end{pmatrix}_{(k-1)\times(k-1)},$$
	then 
	$$\det (\boldsymbol{D}_{3})= \det(\boldsymbol{M}_{2\times 2}) \prod\limits_{t=1}^{k-3}\beta_{i_{t}}\cdot\prod\limits_{1 \leq j<l\leq k-3}\left(\beta _{i_l}-\beta _{i_j}\right).$$
	Note that $\boldsymbol{M}_{2\times 2}\in\mathrm{GL}_{2}(\mathbb{F}_{q})$ and $\beta _{i_t}\in\mathbb{F}_{q}^{*}$ with $\beta _{i_j}\neq \beta _{i_l}(j\neq l)$, 
	hence, the $k-1$ columns of $\boldsymbol{B}_{3}$ are $\mathbb{F}_{q}$-linearly independent.
	
	From the above discussions, we complete the proof of Lemma $\ref{C1dualdistance}$. 
	
	$\hfill\Box$
	
\begin{lemma}\label{noexistcodewords}
For any  $\boldsymbol{c}=(c_{1},\ldots,c_{q+1})\in\mathrm{RL}_{k}^{\perp}\left(\mathbb{F}_{q}^{*},\boldsymbol{M}_{2\times 2}\right)$ with Hamming weight $k$, the following statements are true,

(1) there does not exist any $k$-elements subset $\left\{c_{i_1},\ldots, c_{i_{k}}\right\}$ of $\left\{c_{1},\ldots, c_{q-1}\right\}$ such that  $\prod\limits_{j=1}^{k}c_{i_{j}}\neq 0$ and $c_{q}=c_{q+1}=0$;

(2) there does not exist any $(k-2)$-elements subset $\left\{c_{i_1},\ldots, c_{i_{k-2}}\right\}$ of $\left\{c_{1},\ldots, c_{q-1}\right\}$ such that  $\prod\limits_{j=1}^{k-2}c_{i_{j}}\neq 0$ and $c_{q}c_{q+1}\neq 0$.
\end{lemma}	
{\bf Proof}. For Lemma \ref{noexistcodewords} (1), it's enough to prove that there does not exist any codeword with Hamming weight $k$ in $\mathrm{RL}_{k}^{\perp}\left(\mathbb{F}_{q}^{*},\boldsymbol{M}_{2\times 2}\right)$ as the following form
\begin{equation}\label{nokweightcodeword1}
(\underbrace{0,\ldots,0,c_{i_1},0,\ldots,0,c_{i_2},\ldots,0,\ldots,0,c_{i_{k-1}},0,\ldots,0,c_{i_k},0,\ldots}_{q-1},0,0).
\end{equation}
Otherwise,  if there exists some codeword $\boldsymbol{c}_{11}^{\perp}\in\mathrm{RL}_{k}^{\perp}\left(\mathbb{F}_{q}^{*},\boldsymbol{M}_{2\times 2}\right)$ with Hamming weight $k$ as the form $(\ref{nokweightcodeword1})$,
then $\boldsymbol{c}_{11}^{\perp}\cdot \boldsymbol{G}_1^T=\boldsymbol{0}_{k}$, 
which means that the following homogeneous system of the equations
\begin{equation}\label{lemmaequation1}
	\boldsymbol{F}_{1}\boldsymbol{x}=\mathbf{0}_{k}
\end{equation}
have non-zero solutions, where 
$$\boldsymbol{F}_{1}=
\begin{pmatrix}
	1&\cdots&1\\
	\beta_{i_1}  & \cdots & \beta_{i_{k}}\\
	\vdots&\ddots&\vdots\\
	\beta_{i_1}^{k-3}  & \cdots & \beta_{i_{k}}^{k-3} \\
	\beta_{i_1}^{k-2}  & \cdots & \beta_{i_{k}}^{k-2}\\
	\beta_{i_1}^{k-1} & \cdots & \beta_{i_{k}}^{k-1}  \\
\end{pmatrix}_{k\times k}.
$$
In fact, it's easy to show that
$$\det(\boldsymbol{F}_{1})=\prod_{1\leq j<l\leq k}(\beta_{i_l}-\beta_{i_j})\neq 0,$$	
which means that  $(\ref{lemmaequation1})$ has only zero solution, thus is a contradiction. Thus, Lemma \ref{noexistcodewords} (1) is true.

For Lemma \ref{noexistcodewords} (2), it's enough to prove that there does not exist any codeword with Hamming weight $k$ in $\mathrm{RL}_{k}^{\perp}\left(\mathbb{F}_{q}^{*},\boldsymbol{M}_{2\times 2}\right)$ as the following form
\begin{equation}\label{nokweightcodeword2}
(\underbrace{0,\ldots,0,c_{i_1},0,\ldots,0,c_{i_2},\ldots,0,\ldots,0,c_{i_{k-3}},0,\ldots,0,c_{i_{k-2}},0,\ldots}_{q-1},c_{i_{k-1}},c_{i_{k}}),
\end{equation}
Otherwise,  if  there exists some codeword $\boldsymbol{c}_{12}^{\perp}\in\mathrm{RL}_{k}^{\perp}\left(\mathbb{F}_{q}^{*},\boldsymbol{M}_{2\times 2}\right)$ with Hamming weight $k$ as the form $(\ref{nokweightcodeword2})$,
then $\boldsymbol{c}_{12}^{\perp}\cdot \boldsymbol{G}_{1}^{T}=\boldsymbol{0}_{k}$, 
which means that the following homogeneous system of the equations
\begin{equation}\label{lemmaequation2}
	\boldsymbol{F}_{2}\boldsymbol{x}=\mathbf{0}_{k}
\end{equation}
have non-zero solutions, where 
$$\boldsymbol{F}_{2}=\begin{pmatrix} 
	1&\cdots&1&0&0\\
	\beta_{i_1} & \cdots & \beta_{i_{k-2}}&0&0 \\
	\vdots & \ddots & \vdots& \vdots& \vdots \\
	\beta_{i_1}^{k-3} & \cdots & \beta_{i_{k-2}}^{k-3}&0&0 \\
	\beta_{i_1}^{k-3} & \ldots & \beta_{i_{k-2}}^{k-2}&a_{11}&a_{12} \\
	\beta_{i_1}^{k-1} & \cdots & \beta_{i_{k-2}}^{k-1}&a_{21}&a_{22} \\
\end{pmatrix}_{k\times k}.
$$
In fact, it's easy to show that 
$$\det(\boldsymbol{F}_{2})= \det(\boldsymbol{M}_{2\times 2}) \prod_{1\leq j<l\leq k-2}(\beta_{i_l}-\beta_{i_j})\neq 0,$$	
it means that $(\ref{lemmaequation2})$ has only zero solution, which is a contradiction. Thus, Lemma \ref{noexistcodewords} (2)  is true.

From the above discussions, we complete the proof of Lemma $\ref{noexistcodewords}$.

$\hfill\Box$

\begin{lemma}\label{existcodewords}
For any  $\boldsymbol{c}=(c_{1},\ldots,c_{q+1})\in\mathrm{RL}_{k}^{\perp}\left(\mathbb{F}_{q}^{*},\boldsymbol{M}_{2\times 2}\right)$ with Hamming weight $k$, the following statements are true,
	
	$(1)$ if $a_{21}\in\mathbb{F}_{q}$  and $a_{11}\in\mathbb{F}_{q}^{*}$, then there exist $\# N\left(k-1, \frac{a_{21}}{a_{11}},\mathbb{F}_{q}^{*}\right)$ $(k-1)$-elements subset $\left\{c_{i_1},\ldots, c_{i_{k-1}}\right\}$ of $\left\{c_{1},\ldots, c_{q-1}\right\}$ such that $\prod\limits_{j=1}^{k-1}c_{i_{j}} \neq 0$ and $c_{q} \neq 0, c_{q+1} = 0$;
	
	
	$(2)$ if $a_{21}\in\mathbb{F}_{q}^{*}$ and $a_{11}=0$, then  there does not exist any $(k-1)$-elements subset $\left\{c_{i_1},\ldots, c_{i_{k-1}}\right\}$ of $\left\{c_{1},\ldots, c_{q-1}\right\}$ such that $\prod\limits_{j=1}^{k-1}c_{i_{j}} \neq 0$ and $c_{q}\neq 0, c_{q+1}=0$;
	
	$(3)$ if $a_{22}\in\mathbb{F}_{q}$, and $a_{12}\in\mathbb{F}_{q}^{*}$, then there exist $\# N\left(k-1, \frac{a_{22}}{a_{12}},\mathbb{F}_{q}^{*}\right)$  $(k-1)$-elements subset $\left\{c_{i_1},\ldots, c_{i_{k-1}}\right\}$ of $\left\{c_{1},\ldots, c_{q-1}\right\}$ such that $\prod\limits_{j=1}^{k-1}c_{i_{j}} \neq 0$ and $c_{q}=0, c_{q+1}\neq 0$;
	
	
	$(4)$ if $a_{22}\in\mathbb{F}_{q}^{*}$ and $a_{12}=0$, then  there does not exist any $(k-1)$-elements subset $\left\{c_{i_1},\ldots, c_{i_{k-1}}\right\}$ of $\left\{c_{1},\ldots, c_{q-1}\right\}$ such that $\prod\limits_{j=1}^{k-1}c_{i_{j}} \neq 0$ and $c_{q}= 0, c_{q+1}\neq 0$.
\end{lemma}	
{\bf Proof}. (1) For Lemma \ref{existcodewords} (1), when $a_{21}\in\mathbb{F}_{q}$ and $a_{11}\in\mathbb{F}_{q}^{*}$, it's enough to prove that there exists some codeword with Hamming weight $k$ in $\mathrm{RL}_{k}^{\perp}\left(\mathbb{F}_{q}^{*},\boldsymbol{M}_{2\times 2}\right)$ as the following form
$$(\underbrace{0,\ldots,0,c_{i_1},0,\ldots,0,c_{i_2},\ldots,0,\ldots,0,c_{i_{k-2}},0,\ldots,0,c_{i_{k-1}},0,\ldots}_{q-1},c_{i_k},0),$$
i.e., we only need to prove that the following homogeneous system of the equations
\begin{equation}\label{kweightsystem1}
\boldsymbol{F}_{3}\boldsymbol{x}=\mathbf{0}_{k}
\end{equation}
have non-zero solutions, where 
$$\boldsymbol{F}_{3}=
\begin{pmatrix}
	1&\cdots&1&0\\
	\beta_{i_1}  & \cdots & \beta_{i_{k-1}}& 0  \\
	\vdots&\ddots&\vdots&\vdots\\
	\beta_{i_1}^{k-3}  & \cdots & \beta_{i_{k-1}}^{k-3}&0  \\
	\beta_{i_1}^{k-2}  & \cdots & \beta_{i_{k-1}}^{k-2}& a_{11}  \\
	\beta_{i_1}^{k-1} & \cdots & \beta_{i_{k-1}}^{k-1}& a_{21}  \\
\end{pmatrix}_{k\times k}.
$$
In fact, by Lemma \ref{Vandermonde}, we have
$$\det( \boldsymbol{F}_{3}) =\left(a_{21}-a_{11}\sum\limits_{t=1}^{k-1} \beta_{i_t}\right)\cdot\prod\limits_{\substack{1 \leq j < l \leq k-1}} (\beta_{i_l} - \beta_{i_j}).$$ 
Note that $\prod\limits_{1 \leq j< l\leq k-1}\left(\beta_{i_{l}}-\beta_{i_{j}}\right)\neq 0,$ then it's enough to prove that there exists some $(k-1)$-elements subset $\left\{\beta_{i_1},\ldots, \beta_{i_{k-1}}\right\}$ of $\mathbb{F}_{q}^{*}$ such that $a_{21}-a_{11}\sum\limits_{t=1}^{k-1} \beta_{i_t}=0$. Furthermore, by Lemmas $\ref{subsetsumneq0}$-$\ref{subsetsum0}$, we have $\# N\left(k-1, \frac{a_{21}}{a_{11}},\mathbb{F}_{q}^{*}\right)\neq 0$ for $a_{21}\in\mathbb{F}_{q}$ and $a_{11}\in\mathbb{F}_{q}^{*}$, which means that there exist $\# N\left(k-1, \frac{a_{21}}{a_{11}},\mathbb{F}_{q}^{*}\right)$  $(k-1)$-elements subset $\left\{\beta_{i_1},\ldots, \beta_{i_{k-1}}\right\}$ of $\mathbb{F}_{q}^{*}$ such that $\sum\limits_{t=1}^{k-1} \beta_{i_t}= \frac{a_{21}}{a_{11}}$, i.e., $a_{21}-a_{11}\sum\limits_{t=1}^{k-1} \beta_{i_t}=0$, thus we prove Lemma \ref{existcodewords} (1).




(2) For Lemma \ref{existcodewords} (2), when $a_{21}\in\mathbb{F}_{q}^{*}$ and $a_{11}=0$, it's enough to prove that there does not exist any codeword with Hamming weight $k$ in $\mathrm{RL}_{k}^{\perp}\left(\mathbb{F}_{q}^{*},\boldsymbol{M}_{2\times 2}\right)$ as the following form
$$(\underbrace{0,\ldots,0,c_{i_1},0,\ldots,0,c_{i_2},\ldots,0,\ldots,0,c_{i_{k-2}},0,\ldots,0,c_{i_{k-1}},0,\ldots}_{q-1},c_{i_k},0),$$
i.e., we only need to prove that the following homogeneous system of the equations
\begin{equation}\label{lemmaequation3}
	\boldsymbol{F}_{4}\boldsymbol{x}=\mathbf{0}_{k}
\end{equation}
has only zero solution, where 
$$\boldsymbol{F}_{4}=
\begin{pmatrix}
	1&\cdots&1&0\\
	\beta_{i_1}  & \cdots & \beta_{i_{k-1}}& 0  \\
	\vdots&\ddots&\vdots&\vdots\\
	\beta_{i_1}^{k-3}  & \cdots & \beta_{i_{k-1}}^{k-3}&0  \\
	\beta_{i_1}^{k-2}  & \cdots & \beta_{i_{k-1}}^{k-2}& 0\\
	\beta_{i_1}^{k-1} & \cdots & \beta_{i_{k-1}}^{k-1}& a_{21}  \\
\end{pmatrix}_{k\times k}.
$$
While, by Lemma \ref{Vandermonde}, we have
$$\det( \boldsymbol{F}_{4}) =a_{21}\prod\limits_{\substack{1 \leq j < l \leq k-1}} (\beta_{i_l} - \beta_{i_j})\neq 0.$$ 	 
It means that $(\ref{lemmaequation3})$ has only zero solution, thus Lemma \ref{existcodewords} (2) is true.	 

In the similar proofs as those of Lemma \ref{existcodewords} (1) and (2), we immediately get Lemma \ref{existcodewords} (3)-(4), respectively.

From the above discussions, we complete the proof of Lemma $\ref{existcodewords}$.

$\hfill\Box$
\begin{lemma}\label{C1dualweight}
For $4\leq k\leq q-3$  and $p=2$, or $3\leq k\leq q-2$ and $p\neq 2$, we have $d\left(\mathrm{RL}_{k}^{\perp}\left(\mathbb{F}_{q}^{*},\boldsymbol{M}_{2\times 2}\right)\right)=k$, and the total number $A_k^{\bot}$ of codewords with weight $k$ in $\mathrm{RL}_{k}^{\perp}\left(\mathbb{F}_{q}^{*},\boldsymbol{M}_{2\times 2}\right)$ is
$$A_k^{\bot}=\begin{cases}
\frac{2(q-1)}{q}\binom{q-1}{k-1}+(-1)^{k+\lfloor\frac{k-1}{p}\rfloor}\frac{2(q-1)}{q}\binom{\frac{q}{p}-1}{\lfloor\frac{k-1}{p}\rfloor},&\text{for~}a_{11}a_{12}a_{21}a_{22}\neq 0;\\
\frac{2(q-1)}{q}\binom{q-1}{k-1}+(-1)^{k-1+\lfloor\frac{k-1}{p}\rfloor}\frac{q^2-3q+2}{q}\binom{\frac{q}{p}-1}{\lfloor\frac{k-1}{p}\rfloor},&\text{for~}a_{11}a_{12}a_{21}\neq 0\ \text{and}\ a_{22}= 0;\\
&\ \ \ \text{or~} a_{11}a_{12}a_{22}\neq 0\ \text{and}\ a_{21}= 0;\\
\frac{q-1}{q}\binom{q-1}{k-1}+(-1)^{k+\lfloor\frac{k-1}{p}\rfloor}\frac{q-1}{q}\binom{\frac{q}{p}-1}{\lfloor\frac{k-1}{p}\rfloor},&\text{for~}a_{11}a_{21}a_{22}\neq 0\ \text{and}\ a_{12}= 0;\\
&\ \ \ \text{or~}a_{12}a_{21}a_{22}\neq 0\ \text{and}\ a_{11}= 0;\\
\frac{q-1}{q}\binom{q-1}{k-1}+(-1)^{k-1+\lfloor\frac{k-1}{p}\rfloor}\frac{(q-1)^2}{q}\binom{\frac{q}{p}-1}{\lfloor\frac{k-1}{p}\rfloor},&\text{for~}a_{12}a_{21}\neq 0,a_{11}= 0\ \text{and}\ a_{22}= 0;\\
&\ \ \ \text{or~}a_{11}a_{22}\neq 0,a_{12}= 0\ \text{and}\ a_{21}= 0.\\
\end{cases}$$
\end{lemma}
{\bf Proof}. By Lemmas $\ref{C1dualdistance}$-$\ref{existcodewords}$, we have $d\left(\mathrm{RL}_{k}^{\perp}\left(\mathbb{F}_{q}^{*},\boldsymbol{M}_{2\times 2}\right)\right)\geq k$ and there exists some codeword $\boldsymbol{c}\in\mathrm{RL}_{k}^{\perp}\left(\mathbb{F}_{q}^{*},\boldsymbol{M}_{2\times 2}\right)$ with Hamming weight $k$, and then $d\left(\mathrm{RL}_{k}^{\perp}\left(\mathbb{F}_{q}^{*},\boldsymbol{M}_{2\times 2}\right)\right)=k$.

Now, we calculate the number of codewords with weight $k$ in $\mathrm{RL}_{k}^{\perp}\left(\mathbb{F}_{q}^{*},\boldsymbol{M}_{2\times 2}\right)$ as the following seven cases.

\textbf{Case 1.} If $a_{11}a_{21}a_{12}a_{22}\neq 0$, then by the proofs of Lemma \ref{existcodewords} (1) and (3), we know that for any given $(k-1)$-elements subset, both the homogeneous systems of the equations (\ref{kweightsystem1})-(\ref{lemmaequation3}) have $(q-1)$ non-zero solutions. It means that the total number of codewords with weight $k$ in $\mathrm{RL}_{k}^{\perp}\left(\mathbb{F}_{q}^{*},\boldsymbol{M}_{2\times 2}\right)$ is  
$$\begin{aligned}
A_k^{\bot}=&(q-1)\# N\left(k-1, \frac{a_{21}}{a_{11}},\mathbb{F}_{q}^{*}\right)+(q-1)\# N\left(k-1, \frac{a_{22}}{a_{12}},\mathbb{F}_{q}^{*}\right)\\
=&\frac{2(q-1)}{q}\binom{q-1}{k-1}+(-1)^{k+\lfloor\frac{k-1}{p}\rfloor}\frac{2(q-1)}{q}\binom{\frac{q}{p}-1}{\lfloor\frac{k-1}{p}\rfloor}.
\end{aligned}$$	

\textbf{Case 2.} If $a_{11}a_{12}a_{21}\neq 0$ and $a_{22}=0$, then by Lemma \ref{existcodewords} (1) and (3), the total number of codewords with weight $k$ in $\mathrm{RL}_{k}^{\perp}\left(\mathbb{F}_{q}^{*},\boldsymbol{M}_{2\times 2}\right)$ is
$$\begin{aligned}
	A_k^{\bot}=&(q-1)\# N\left(k-1, \frac{a_{21}}{a_{11}},\mathbb{F}_{q}^{*}\right)+(q-1)\# N\left(k-1,0,\mathbb{F}_{q}^{*}\right)\\
	=&\frac{2(q-1)}{q}\binom{q-1}{k-1}+(-1)^{k-1+\lfloor\frac{k-1}{p}\rfloor}\frac{q^2-3q+2}{q}\binom{\frac{q}{p}-1}{\lfloor\frac{k-1}{p}\rfloor}.\\
\end{aligned}$$	

\textbf{Case 3.} If $a_{11}a_{21}a_{22}\neq 0$ and $a_{12}=0$, then by Lemma \ref{existcodewords} (1) and (4), the total number of codewords with weight $k$ in $\mathrm{RL}_{k}^{\perp}\left(\mathbb{F}_{q}^{*},\boldsymbol{M}_{2\times 2}\right)$ is
$$A_k^{\bot}=(q-1)\# N\left(k-1, \frac{a_{21}}{a_{11}},\mathbb{F}_{q}^{*}\right)=\frac{q-1}{q}\binom{q-1}{k-1}+(-1)^{k+\lfloor\frac{k-1}{p}\rfloor}\frac{q-1}{q}\binom{\frac{q}{p}-1}{\lfloor\frac{k-1}{p}\rfloor}.$$	

\textbf{Case 4.} If $a_{11}a_{12}a_{22}\neq 0$ and $a_{21}=0$, then by Lemma \ref{existcodewords} (1) and (3), the total number of codewords with weight $k$ in $\mathrm{RL}_{k}^{\perp}\left(\mathbb{F}_{q}^{*},\boldsymbol{M}_{2\times 2}\right)$ is 
$$\begin{aligned}
	A_k^{\bot}=&(q-1)\# N\left(k-1,0,\mathbb{F}_{q}^{*}\right)+(q-1)\# N\left(k-1, \frac{a_{22}}{a_{12}},\mathbb{F}_{q}^{*}\right)\\
	=&\frac{2(q-1)}{q}\binom{q-1}{k-1}+(-1)^{k-1+\lfloor\frac{k-1}{p}\rfloor}\frac{q^2-3q+2}{q}\binom{\frac{q}{p}-1}{\lfloor\frac{k-1}{p}\rfloor}.
\end{aligned}$$	

\textbf{Case 5.}  If $a_{11}a_{22}\neq 0,a_{21}=0$ and $a_{12}=0$,  then by Lemma \ref{existcodewords} (1) and (4), the total number of codewords with weight $k$ in $\mathrm{RL}_{k}^{\perp}\left(\mathbb{F}_{q}^{*},\boldsymbol{M}_{2\times 2}\right)$ is
$$A_k^{\bot}=(q-1)\# N\left(k-1,0,\mathbb{F}_{q}^{*}\right)=\frac{q-1}{q}\binom{q-1}{k-1}+(-1)^{k-1+\lfloor\frac{k-1}{p}\rfloor}\frac{(q-1)^2}{q}\binom{\frac{q}{p}-1}{\lfloor\frac{k-1}{p}\rfloor}.$$	

\textbf{Case 6.}  If $a_{12}a_{21}a_{22}\neq 0$ and $a_{11}=0$, then by Lemma \ref{existcodewords} (2) and (3), the total number of codewords with weight $k$ in $\mathrm{RL}_{k}^{\perp}\left(\mathbb{F}_{q}^{*},\boldsymbol{M}_{2\times 2}\right)$ is 
$$A_k^{\bot}=(q-1)\# N\left(k-1, \frac{a_{22}}{a_{12}},\mathbb{F}_{q}^{*}\right)=\frac{q-1}{q}\binom{q-1}{k-1}+(-1)^{k+\lfloor\frac{k-1}{p}\rfloor}\frac{q-1}{q}\binom{\frac{q}{p}-1}{\lfloor\frac{k-1}{p}\rfloor}.$$

\textbf{Case 7.} If $a_{12}a_{21}\neq 0,a_{11}=0$ and $a_{22}=0$, then by Lemma \ref{existcodewords} (2) and (3), the total number of codewords with weight $k$ in $\mathrm{RL}_{k}^{\perp}\left(\mathbb{F}_{q}^{*},\boldsymbol{M}_{2\times 2}\right)$ is
$$A_k^{\bot}=(q-1)\# N\left(k-1,0,\mathbb{F}_{q}^{*}\right)=\frac{q-1}{q}\binom{q-1}{k-1}+(-1)^{k-1+\lfloor\frac{k-1}{p}\rfloor}\frac{(q-1)^2}{q}\binom{\frac{q}{p}-1}{\lfloor\frac{k-1}{p}\rfloor}.$$

From the above discussions, we complete the proof of Lemma $\ref{C1dualweight}$.

$\hfill\Box$

\begin{theorem}\label{RL1}
For $4\leq k\leq q-3$  and $p=2$, or $3\leq k\leq q-2$ and $p\neq 2$, the code $\mathrm{RL}_{k}\left(\mathbb{F}_{q}^{*},\boldsymbol{M}_{2\times 2}\right)$ is NMDS with the parameters $[q+1,k,q+1-k]_q$, and the total number  $A_{q+1-k}^{\bot}$  of codewords with weight $q+1-k$ in $\mathrm{RL}_{k}\left(\mathbb{F}_{q}^{*},\boldsymbol{M}_{2\times 2}\right)$ is
$$A_{q+1-k}=\begin{cases}
	\frac{2(q-1)}{q}\binom{q-1}{k-1}+(-1)^{k+\lfloor\frac{k-1}{p}\rfloor}\frac{2(q-1)}{q}\binom{\frac{q}{p}-1}{\lfloor\frac{k-1}{p}\rfloor},&\text{for~}a_{11}a_{12}a_{21}a_{22}\neq 0;\\
	\frac{2(q-1)}{q}\binom{q-1}{k-1}+(-1)^{k-1+\lfloor\frac{k-1}{p}\rfloor}\frac{q^2-3q+2}{q}\binom{\frac{q}{p}-1}{\lfloor\frac{k-1}{p}\rfloor},&\text{for~}a_{11}a_{12}a_{21}\neq 0\ \text{and}\ a_{22}= 0;\\
	&\ \ \ \text{or~} a_{11}a_{12}a_{22}\neq 0\ \text{and}\ a_{21}= 0;\\
	\frac{q-1}{q}\binom{q-1}{k-1}+(-1)^{k+\lfloor\frac{k-1}{p}\rfloor}\frac{q-1}{q}\binom{\frac{q}{p}-1}{\lfloor\frac{k-1}{p}\rfloor},&\text{for~}a_{11}a_{21}a_{22}\neq 0\ \text{and}\ a_{12}= 0;\\
	&\ \ \ \text{or~}a_{12}a_{21}a_{22}\neq 0\ \text{and}\ a_{11}= 0;\\
	\frac{q-1}{q}\binom{q-1}{k-1}+(-1)^{k-1+\lfloor\frac{k-1}{p}\rfloor}\frac{(q-1)^2}{q}\binom{\frac{q}{p}-1}{\lfloor\frac{k-1}{p}\rfloor},&\text{for~}a_{12}a_{21}\neq 0,a_{11}= 0\ \text{and}\ a_{22}= 0;\\
	&\ \ \ \text{or~}a_{11}a_{22}\neq 0,a_{12}= 0\ \text{and}\ a_{21}= 0.\\
\end{cases}$$
\end{theorem}
{\bf Proof}. It's easy to prove that $\rank(\boldsymbol{G}_1)=k$, hence, ${\rm dim}(\mathrm{RL}_{k}\left(\mathbb{F}_{q}^{*},\boldsymbol{M}_{2\times 2}\right))=\rank(\boldsymbol{G}_1)=k$, which means that $\dim{(\mathrm{RL}_{k}^{\perp}\left(\mathbb{F}_{q}^{*},\boldsymbol{M}_{2\times 2}\right))}=q+1-\dim{(\mathrm{RL}_{k}\left(\mathbb{F}_{q}^{*},\boldsymbol{M}_{2\times 2}\right))}=q+1-k$, furthermore, by Lemma \ref{C1dualweight}, $\mathrm{RL}_{k}^{\perp}\left(\mathbb{F}_{q}^{*},\boldsymbol{M}_{2\times 2}\right)$ is AMDS with the parameters $[q+1,q+1-k,k]_q$.

Next, we prove that $d(\mathrm{RL}_{k}\left(\mathbb{F}_{q}^{*},\boldsymbol{M}_{2\times 2}\right))=q+1-k$. Assume that $d(\mathrm{RL}_{k}\left(\mathbb{F}_{q}^{*},\boldsymbol{M}_{2\times 2}\right))\leq q-k$. Then, for any nonzero codeword $\boldsymbol{c}$ with minimum weight $d$ in $\mathrm{RL}_{k}\left(\mathbb{F}_{q}^{*},\boldsymbol{M}_{2\times 2}\right)$, there are at least $k+1$ zero coordinates. Let $\boldsymbol{g}_i$ be the $(i+1)$-th row of $\boldsymbol{G}_1$, thus we can assume that 
$$
\boldsymbol{c}=b_0\boldsymbol{g}_0+\cdots+b_{k-1}\boldsymbol{g}_{k-1}=\begin{pmatrix}
	b_{0}+b_{1}\beta_{1}+\cdots+b_{k-1}\beta_{1}^{k-1}\\
	\vdots\\
	b_{0}+b_{1}\beta_{q-1}+\cdots+b_{k-1}\beta_{q-1}^{k-1}\\
	
	b_{k-2}a_{11}+b_{k-1}a_{21}\\
	b_{k-2}a_{12}+b_{k-1}a_{22}
\end{pmatrix}_{(q+1)\times 1}^T=\left(c_{1},\ldots,c_{q-1},c_{q},c_{q+1}\right),
$$ 
where $b_j\in\mathbb{F}_{q}(j=0,\ldots,k-1)$, and so we have the following three cases .

\textbf{Case 1.} If there exists some codeword $\boldsymbol{c}=(c_{1},\ldots,c_{q-1},c_{q},c_{q+1})$ with $c_{i_1}=\ldots=c_{i_{k-1}}=c_{q}=c_{q+1}=0$ , where $1\leq i_{j}\leq q-1(j=1,\ldots,k-1)$, it means that there exists some $(k-1)$-elements subset $\left\{\beta_{i_1},\ldots, \beta_{i_{k-1}}\right\}$ of $\mathbb{F}_{q}^{*}$ such that
\begin{equation}\label{case11}
b_{0}+b_{1}\beta_{i}+\cdots+b_{k-1}\beta_{i}^{k-1}=0\ ( i=i_1,\ldots,i_{k-1}),
\end{equation}
and
\begin{equation}\label{case12}
	\begin{cases}
		b_{k-2}a_{11}+b_{k-1}a_{21}=0,\\
		b_{k-2}a_{12}+b_{k-1}a_{22}=0.
	\end{cases}
\end{equation}
Note that $\det \begin{pmatrix}
a_{11}&a_{21}\\
a_{12}&a_{22}
\end{pmatrix}=a_{11}a_{22}-
a_{21}a_{12}=\det(\boldsymbol{M}_{2\times 2}^{T})\neq 0,$ it means that the homegeneous systen of the equations (\ref{case12}) has only zero solution, i.e., $b_{k-2}=b_{k-1}=0.$ Furthermore, by (\ref{case11}) we know that for any $i\in\{i_1,\ldots,i_{k-1}\},$
$$b_{0}+b_{1}\beta_{i}+\cdots+b_{k-3}\beta_{i}^{k-3}=0,$$
i.e., the polynomial $f_{1}(x)=b_{0}+b_{1}x+\cdots+b_{k-3}x^{k-3}$ has $k-1$ roots over $\mathbb{F}_{q}$, 
which is a contradiction with $\deg(f(x))=k-3$.

\textbf{Case 2.} If there exists some codeword $\boldsymbol{c}=(c_{1},\ldots,c_{q-1},c_{q},c_{q+1})$ such that $c_{i_1}=\ldots=c_{i_{k}}=0$ and one of $c_{q}$ and $c_{q+1}$ is zero, where $1\leq i_{j}\leq q-1(j=1,\ldots,k)$, it means that there exists some  $k$-elements subset $\left\{\beta_{i_1},\ldots, \beta_{i_{k}}\right\}$ of $\mathbb{F}_{q}^{*}$ such that
\begin{equation}\label{case21}
	\begin{cases}
		b_{0}+b_{1}\beta_{i}+\cdots+b_{k-1}\beta_{i}^{k-1}=0(i=i_1,\ldots,i_{k}), \\
		b_{k-2}a_{11}+b_{k-1}a_{21}=0.
	\end{cases}
\end{equation}
or
\begin{equation}\label{case22}
	\begin{cases}
		b_{0}+b_{1}\beta_{i}+\cdots+b_{k-1}\beta_{i}^{k-1}=0(i=i_1,\ldots,i_{k}), \\
		b_{k-2}a_{12}+b_{k-1}a_{22}=0.
	\end{cases}
\end{equation}
Furthermore, by $(\ref{case21})$-$(\ref{case22})$, we know that the polynomial $f_{2}(x)=b_{0}+b_{1}x+\cdots+b_{k-1}x^{k-1}$ has $k$ roots over $\mathbb{F}_{q}$, 
which is a contradiction with $\deg(f(x))=k-1$.

\textbf{Case 3.} If there exists some codeword $\boldsymbol{c}=(c_{1},\ldots,c_{q-1},c_{q},c_{q+1})$ with $c_{i_1}=\ldots=c_{i_{k+1}}=0$, where $1\leq i_{j}\leq q-1(j=1,\ldots,k+1)$, it means that there exists some $(k+1)$-elements subset $\left\{\beta_{i_1},\ldots, \beta_{i_{k+1}}\right\}$ of $\mathbb{F}_{q}^{*}$ such that
$$b_{0}+b_{1}\beta_{i}+\cdots+b_{k-1}\beta_{i}^{k-1}=0\ (i=i_1,\ldots,i_{k+1}),$$
it's means that the polynomial $f_{3}(x)=b_{0}+b_{1}x+\cdots+b_{k-1}x^{k-1}$ has $k+1$ roots over $\mathbb{F}_{q}$, 
which is a contradiction with $\deg(f(x))=k-1$.

From the above discussions, we have $d(\mathrm{RL}_{k}\left(\mathbb{F}_{q}^{*},\boldsymbol{M}_{2\times 2}\right))\geq q+1-k$. Note that by the Singleton bound, we have $d(\mathrm{RL}_{k}\left(\mathbb{F}_{q}^{*},\boldsymbol{M}_{2\times 2}\right)) \leq q+2 - k $, then
$$q+1-k\leq d(\mathrm{RL}_{k}\left(\mathbb{F}_{q}^{*},\boldsymbol{M}_{2\times 2}\right)) \leq q+2 - k.$$
If $d(\mathrm{RL}_{k}\left(\mathbb{F}_{q}^{*},\boldsymbol{M}_{2\times 2}\right))=q+2-k$, then $\mathrm{RL}_{k}\left(\mathbb{F}_{q}^{*},\boldsymbol{M}_{2\times 2}\right)$ is MDS with the parameters $[q+1,k,q+2-k]_q$, and so by Lemma \ref{MDSdefinition},  $\mathrm{RL}_{k}^{\perp}\left(\mathbb{F}_{q}^{*},\boldsymbol{M}_{2\times 2}\right)$ is also  MDS, which contradicts with the fact  that $\mathrm{RL}_{k}^{\perp}\left(\mathbb{F}_{q}^{*},\boldsymbol{M}_{2\times 2}\right)$ is AMDS with the parameters $[q+1,q+1-k,k]_q$. Hence, $d(\mathrm{RL}_{k}\left(\mathbb{F}_{q}^{*},\boldsymbol{M}_{2\times 2}\right))=q+1-k$, it's mean that  $\mathrm{RL}_{k}\left(\mathbb{F}_{q}^{*},\boldsymbol{M}_{2\times 2}\right)$ is NMDS with the parameters $[q+1,k,q+1-k]_q$, and then by Lemma  \ref{NMDSweight}, $A_{q+1-k}=A_{k}^{\perp}$.

This completes the proof of Theorem $\ref{RL1}$.

$\hfill\Box$

By the proofs of Lemmas $\ref{noexistcodewords}$-$\ref{existcodewords}$ and Theorem $\ref{RL1}$, we can immediately obtain the following 
\begin{corollary}\label{Fq0MDS}
If $k\in\left\{3,q-2,q-1\right\}$ and $p=2$, or $q-1$ and $p\neq 2$, then $\mathrm{RL}_{k}\left(\mathbb{F}_{q}^{*},\boldsymbol{M}_{2\times 2}\right)$ is MDS with the parameters $[q+1,k,q+2-k]_q$.
\end{corollary}

The following Example \ref{RL1example} is for Theorem $\ref{RL1}$ in the case $\boldsymbol{M}_{2\times 2}=\begin{pmatrix}
	1 & 1 \\
	2 & 1 \\
\end{pmatrix}\in\mathrm{GL}_{2}(\mathbb{F}_{q}).$
\begin{example}\label{RL1example}
Let $q=3^m$ and $k=5$, then 
$$\boldsymbol{G}_1=
\begin{pmatrix}
1 &          \cdots & 1              & 0 & 0  \\
\beta_1   &  \cdots & \beta_{q-1}   & 0 & 0 \\
\beta_1^2 &  \cdots & \beta_{q-1}^2 & 0 & 0 \\
\beta_1^3 &  \cdots & \beta_{q-1}^3 & 1 & 1 \\
\beta_1^4 &  \cdots & \beta_{q-1}^4 & 2 & 1 \\
\end{pmatrix}.
$$
By Theorem \ref{RL1}, $\mathrm{RL}_{k}\left(\mathbb{F}_{q}^{*},\boldsymbol{M}_{2\times 2}\right)$ is NMDS with the parameters $[q+1,5,q-4]_{q}$ and 
$$A_{q-4}=A_5^\perp=(q-1)\# N(4,2,\mathbb{F}_{q}^{*})+(q-1)\# N(4,1,\mathbb{F}_{q}^{*})=\frac{(q-1)(q-3)(q^2-7q+14)}{12}.$$ 
Furthermore, by Lemma \ref{NMDSweight}, the weight enumerator of $\mathrm{RL}_{k}\left(\mathbb{F}_{q}^{*},\boldsymbol{M}_{2\times 2}\right)$ is	
$$
\begin{aligned}
A(x)=&1+A_{q-4}x^{q-4}+A_{q-3}x^{q-3}+A_{q-2}x^{q-2}+A_{q-1}x^{q-1}+A_{q}x^{q}+A_{q+1}x^{q+1}\\
=&1+\frac{(q-1)(q-3)(q^2-7q+14)}{12}x^{q-4}+\frac{(q-1)\left(q^4-12q^3+99q^2-348q+420\right)}{24}x^{q-3}\\
&+\frac{(q-1)\left(16q^3-100q^2+344q-420\right)}{12}x^{q-2}+\frac{(q-1)\left(3q^4-10q^3+115q^2-332q+420\right)}{12}x^{q-1}\\
&+\frac{(q-1)\left(4q^4+15q^3-54q^2+177q-198\right)}{12}x^{q}+\frac{(q-1)\left(9q^4-8q^3+23q^2-76q+84\right)}{24}x^{q+1}.
\end{aligned}
$$
Let $m=2,\mathbb{F}_{3^2}^{*}=\langle\omega\rangle$ and $\beta_{i}=\omega^{i-1}(i=1,\ldots,8)$, then
$$\boldsymbol{G}_1=
\begin{pmatrix}
	1 &1&1&1&1&1 & 1&1& 0 & 0  \\
	1   &\omega&\omega^{2}&\omega^{3}&2&\omega^{5}&\omega^{6}&\omega^{7}& 0 & 0 \\
	1 &\omega^{2}&2&\omega^{6}&1&\omega^{2}&2&\omega^{6} & 0 & 0 \\
	1 &\omega^{3}&\omega^{6}&\omega&2&\omega^{7}&\omega^{2}&\omega^{5} & 1 & 1 \\
		1 &2&1&2&1&2&1&2& 2 & 1 \\
\end{pmatrix}.
$$
Based on the Magma programe, $\mathrm{RL}_{k}\left(\mathbb{F}_{q}^{*},\boldsymbol{M}_{2\times 2}\right)$ is NMDS with the parameters $[10,5,5]_{3^2}$ and its weight enumerator is
$$A(x)=1+128x^5+1040x^6+4160x^7+12760x^8+22800x^9+18160x^{10},$$
which is consistent with Theorem \ref{C1dualweight}.
\end{example}

\section{The NMDS property of $\text{RL}_{k}\left(\mathbb{F}_{q},M_{2\times 2}\right)$}

Let $\mathbb{F}_{q}=\left\{\alpha_1,\ldots,\alpha_{q}\right\}$, in this section, we prove that  $\mathrm{RL}_{k}\left(\mathbb{F}_{q},\boldsymbol{M}_{2\times 2}\right)$ is NMDS,
and determine its weight distribution for both $4\leq k\leq q-2$ and $p=2$, or  both $3\leq k\leq q$ and $p\neq 2$,  where Char($\mathbb{F}_{q}$)=$p$. 
\subsection{Main results}
In this subsection, by the similar proofs as those for Lemmas \ref{C1dualdistance}-\ref{existcodewords} and Theorem \ref{C1dualweight}, one can
obtain the following Lemma \ref{C3dualweight} and Theorem \ref{C3dualweight}.

\begin{lemma}\label{C3dualweight}
For $4\leq k\leq q-2$ and $p=2$, or  $3\leq k\leq q$ and $p\neq 2$, we have $d\left(\mathrm{RL}_{k}^{\perp}\left(\mathbb{F}_{q},\boldsymbol{M}_{2\times 2}\right)\right)=k$, and for the total number $A_k^{\bot}$ of codewords with Hamming weight $k$ in $\mathrm{RL}_{k}^{\perp}\left(\mathbb{F}_{q},\boldsymbol{M}_{2\times 2}\right)$, the following statements are true,

$(1)$ if $p\nmid k-1$, then
	$$ A_k^{\bot}=\begin{cases}
		 \frac{2(q-1)}{q}\binom{q}{k-1},&\text{for~}a_{11}a_{12}a_{21}a_{22}\neq 0;\\
		 &\ \ \ \text{or~}a_{11}a_{12}a_{21}\neq 0\ \text{and~}a_{22}=0;\\
		 &\ \ \ \text{or~} a_{11}a_{12}a_{22}\neq 0\ \text{and}\ a_{21}=0;\\
		 \frac{q-1}{q}\binom{q}{k-1},&\text{for~}a_{11}a_{21}a_{22}\neq  0\ \text{and~}a_{12}=0;\\
		 &\ \ \ \text{or~}a_{12}a_{21}a_{22}\neq  0\ \text{and~}a_{11}=0;\\
		 &\ \ \ \text{or~}a_{12}a_{21}\neq 0, a_{11}=0\ \text{and~}a_{22}=0;\\
		 &\ \ \ \text{or~} a_{11}a_{22}\neq 0,a_{12}=0\ \text{and~}a_{21}=0.\\
	\end{cases}$$
	
$(2)$ if $p\mid k-1$, then
$$ A_k^{\bot}=\begin{cases}
	\frac{2(q-1)}{q}\binom{q}{k-1}+(-1)^{k+\frac{k-1}{p}}\frac{2(q-1)}{q}\binom{\frac{q}{p}}{\frac{k-1}{p}},&\text{for~}a_{11}a_{21}a_{12}a_{22}\neq 0;\\
	\frac{2(q-1)}{q}\binom{q}{k-1}+(-1)^{k-1+\frac{k-1}{p}}\frac{q^2-3q+2}{q}\binom{\frac{q}{p}}{\frac{k-1}{p}},&\text{for~}a_{11}a_{12}a_{21}\neq 0\ \text{and~}a_{22}= 0;\\
	 &\ \ \ \ \text{or~} a_{11}a_{12}a_{22}\neq 0\ \text{and~}a_{21}= 0;\\
	\frac{q-1}{q}\binom{q}{k-1}+(-1)^{k+\frac{k-1}{p}}\frac{q-1}{q}\binom{\frac{q}{p}}{\frac{k-1}{p}},&\text{for~}a_{11}a_{21}a_{22}\neq 0\ \text{and~}a_{12}= 0;\\
	\frac{q-1}{q}\binom{q}{k-1}+(-1)^{k+\frac{k-1}{p}}\frac{q-1}{q}\binom{\frac{q}{p}}{\frac{k-1}{p}},&\text{for~}a_{12}a_{21}a_{22}\neq  0\ \text{and~}a_{11}= 0;\\
		
		\frac{q-1}{q}\binom{q}{k-1}+(-1)^{k-1+\frac{k-1}{p}}\frac{(q-1)^2}{q}\binom{\frac{q}{p}}{\frac{k-1}{p}},&\text{for~}a_{12}a_{21}\neq 0, a_{11}= 0\ \text{and~}a_{22}= 0;\\ 
		&\ \ \ \  \text{or~} a_{22}a_{11}\neq 0,a_{21}=0\ \text{and~}a_{12}=0.\\
\end{cases}$$
\end{lemma}
\begin{theorem}\label{RL2}
For $4\leq k\leq q-2$ and $p=2$, or  $3\leq k\leq q$ and $p\neq 2$, the code $\mathrm{RL}_{k}\left(\mathbb{F}_{q},\boldsymbol{M}_{2\times 2}\right)$ is NMDS with the parameters $[q+2,k,q+2-k]_q$, and for the total number $A_{q+2-k}^{\bot}$ of codewords with Hamming weight $q+2-k$ in $\mathrm{RL}_{k}\left(\mathbb{F}_{q},\boldsymbol{M}_{2\times 2}\right)$, the following statements are true,

$(1)$ if $p\nmid k-1$, then
$$ A_{q+2-k}^{\bot}=\begin{cases}
	\frac{2(q-1)}{q}\binom{q}{k-1},&\text{for~}a_{11}a_{12}a_{21}a_{22}\neq 0;\\
	&\ \ \ \text{or~}a_{11}a_{12}a_{21}\neq 0\ \text{and~}a_{22}=0;\\
	&\ \ \ \text{or~} a_{11}a_{12}a_{22}\neq 0\ \text{and}\ a_{21}=0;\\
	\frac{q-1}{q}\binom{q}{k-1},&\text{for~}a_{11}a_{21}a_{22}\neq  0\ \text{and~}a_{12}=0;\\
	&\ \ \ \text{or~}a_{12}a_{21}a_{22}\neq  0\ \text{and~}a_{11}=0;\\
	&\ \ \ \text{or~}a_{12}a_{21}\neq 0, a_{11}=0\ \text{and~}a_{22}=0;\\
	&\ \ \ \text{or~} a_{11}a_{22}\neq 0,a_{12}=0\ \text{and~}a_{21}=0;\\
\end{cases}$$

$(2)$ if $p\mid k-1$, then
$$ A_{q+2-k}^{\bot}=\begin{cases}
	\frac{2(q-1)}{q}\binom{q}{k-1}+(-1)^{k+\frac{k-1}{p}}\frac{2(q-1)}{q}\binom{\frac{q}{p}}{\frac{k-1}{p}},&\text{for~}a_{11}a_{21}a_{12}a_{22}\neq 0;\\
	\frac{2(q-1)}{q}\binom{q}{k-1}+(-1)^{k-1+\frac{k-1}{p}}\frac{q^2-3q+2}{q}\binom{\frac{q}{p}}{\frac{k-1}{p}},&\text{for~}a_{11}a_{12}a_{21}\neq 0\ \text{and~}a_{22}= 0;\\
	&\ \ \ \ \text{or~} a_{11}a_{12}a_{22}\neq 0\ \text{and~}a_{21}= 0;\\
	\frac{q-1}{q}\binom{q}{k-1}+(-1)^{k+\frac{k-1}{p}}\frac{q-1}{q}\binom{\frac{q}{p}}{\frac{k-1}{p}},&\text{for~}a_{11}a_{21}a_{22}\neq 0\ \text{and~}a_{12}= 0;\\
	\frac{q-1}{q}\binom{q}{k-1}+(-1)^{k+\frac{k-1}{p}}\frac{q-1}{q}\binom{\frac{q}{p}}{\frac{k-1}{p}},&\text{for~}a_{12}a_{21}a_{22}\neq  0\ \text{and~}a_{11}= 0;\\
	
	\frac{q-1}{q}\binom{q}{k-1}+(-1)^{k-1+\frac{k-1}{p}}\frac{(q-1)^2}{q}\binom{\frac{q}{p}}{\frac{k-1}{p}},&\text{for~}a_{12}a_{21}\neq 0, a_{11}= 0\ \text{and~}a_{22}= 0;\\ 
	&\ \ \ \  \text{or~} a_{22}a_{11}\neq 0,a_{21}=0\ \text{and~}a_{12}=0.\\
\end{cases}$$
\end{theorem}

The following Examples \ref{RL2example3}-\ref{RL2example4} are for Theorem \ref{RL2} in the case $\boldsymbol{M}_{2\times 2}=\begin{pmatrix} 
	1 & 1 \\            
	2 & 1 \\
\end{pmatrix}\in\mathrm{GL}_{2}(\mathbb{F}_{q}).$

\begin{example}\label{RL2example3}
	Let $q=3^m$ and $k=4$, then	$$\boldsymbol{G}_{3}=
	\begin{pmatrix}
		1 &          \cdots & 1 & 0 & 0  \\
		\alpha_1 &  \cdots & \alpha_{q}& 0 & 0 \\
		\alpha_1^2 &  \cdots & \alpha_{q}^2& 1 & 1\\
		\alpha_1^3 &  \cdots & \alpha_{q}^3& 2 & 1 
	\end{pmatrix}.
	$$
	By Theorem \ref{RL2}, $\mathrm{RL}_{k}\left(\mathbb{F}_{q},\boldsymbol{M}_{2\times 2}\right)$ is NMDS with the parameters $[q+2,4,q-2]_{q}$ and 
	$$A_{q-2}=A_4^\perp=(q-1)\# N(3,2,\mathbb{F}_{q})+(q-1)\# N(3,1,\mathbb{F}_{q})=\frac{q(q-1)(q-3)}{3}.$$ Furthermore,  by Lemma \ref{NMDSweight}, the weight enumerator of $\mathrm{RL}_{k}\left(\mathbb{F}_{q},\boldsymbol{M}_{2\times 2}\right)$ is	
	$$
	\begin{aligned}
		A(x)=&1+A_{q-2}x^{q-2}+A_{q-1}x^{q-1}+A_{q}x^{q}+A_{q+1}x^{q+1}+A_{q+2}x^{q+2}\\
		=&1+\frac{q(q-1)(q-3)}{3}x^{q-2}+\frac{q(q-1)(q^2-5q+26)}{6}x^{q-1}+\frac{(q-1)(5q^2-9q+2)}{2}x^{q}\\
		&+\frac{q(q-1)(3q^2-5q+18)}{6}x^{q+1}+\frac{q(q-1)(2q^2-q-5)}{6}x^{q+2}.
	\end{aligned}
	$$
	
	Let $m=2,\mathbb{F}_{3^2}^{*}=\langle\omega\rangle, \alpha_{1}=0$ and $\alpha_{i}=\omega^{i-2}(i=2,\ldots,9)$, then
	$$\begin{pmatrix}
		1&1 &1&1&1&1&1 & 1&1& 0 & 0  \\
		0&1   &\omega&\omega^{2}&\omega^{3}&2&\omega^{5}&\omega^{6}&\omega^{7}&  0 & 0 \\
		0&1 &\omega^{2}&2&\omega^{6}&1&\omega^{2}&2&\omega^{6} &  1 & 1 \\
		0&1 &\omega^{3}&\omega^{6}&\omega&2&\omega^{7}&\omega^{2}&\omega^{5} & 2 & 1
	\end{pmatrix}.
	$$
	Based on the Magma programe, $\mathrm{RL}_{k}\left(\mathbb{F}_{q} ,\boldsymbol{M}_{2\times 2}\right)$ is NMDS with the parameters $[11,4,7]_{3^2}$ and its weight enumerator is
	$$A(x)=1+144x^7+744x^8+1304x^9+2592x^{10}+1776x^{11},$$
	which is consistent with Theorem \ref{RL2}.
\end{example}	

\begin{example}\label{RL2example4}
	Let $q=3^m$ and $k=6$, then
	$$\boldsymbol{G}_{3}=
	\begin{pmatrix}
		1 &          \cdots & 1 & 0 & 0  \\
		\alpha_1   &  \cdots & \alpha_{q}& 0 & 0 \\
		\alpha_1^2 &  \cdots & \alpha_{q}^2& 0 & 0 \\
		\alpha_1^3 &  \cdots & \alpha_{q}^3& 0 & 0 \\
		\alpha_1^4 &  \cdots & \alpha_{q}^4& 1 & 1 \\
		\alpha_1^5 &  \cdots & \alpha_{q}^5& 2 & 1 \\
	\end{pmatrix}.
	$$
	By Theorem \ref{RL2}, $\mathrm{RL}_{k}\left(\mathbb{F}_{q},\boldsymbol{M}_{2\times 2}\right)$ is NMDS with the parameters $[q+2,6,q-4]_{q}$ and 
	$$A_{q-4}=A_{6}^\perp=(q-1)\# N(5,2,\mathbb{F}_{q})+(q-1)\# N(5,1,\mathbb{F}_{q})=\frac{(q-1)^2(q-2)(q-3)(q-4)}{60}.$$ Furthermore, by Lemma \ref{NMDSweight}, the weight enumerator of $\mathrm{RL}_{k}\left(\mathbb{F}_{q},\boldsymbol{M}_{2\times 2}\right)$ is	
	$$
	\begin{aligned}
		A(x)=&1+A_{q-4}x^{q-4}+A_{q-3}x^{q-3}+A_{q-2}x^{q-2}+A_{q-1}x^{q-1}+A_{q}x^{q}+A_{q+1}x^{q+1}+A_{q+2}x^{q+2}\\
		=&1+\frac{(q-1)^2(q-2)(q-3)(q-4)}{60}x^{q-4}+\frac{(q-1)^2(q-2)(q^3-9q^2+86q-144)}{120}x^{q-3}\\
		&+\frac{(q-1)^2(3q^3-15q^2+54q-48)}{8}x^{q-2}+\frac{(q-1)(q^5-2q^4+45q^3-124q^2+212q-96)}{12}x^{q-1}\\
		&+\frac{(q-1)(q^5+6q^4-11q^3+51q^2-71q+42)}{6}x^{q}\\
		&+\frac{(q-1)(15q^5+16q^4+25q^3-140q^2+180q-96)}{40}x^{q+1}\\
		&+\frac{(q-1)(44q^5-63q^4-10q^3+75q^2-94q+48)}{120}x^{q+2}.
	\end{aligned}
	$$
	
	Let $m=2,\mathbb{F}_{3^2}^{*}=\langle\omega\rangle, \alpha_{1}=0$ and $\alpha_{i}=\omega^{i-2}(i=2,\ldots,9)$, then
	$$\begin{pmatrix}
		1&1 &1&1&1&1&1 & 1&1& 0 & 0  \\
		0&1   &\omega&\omega^{2}&\omega^{3}&2&\omega^{5}&\omega^{6}&\omega^{7}&  0 & 0 \\
		0&1 &\omega^{2}&2&\omega^{6}&1&\omega^{2}&2&\omega^{6} &  0 & 0 \\
		0&1 &\omega^{3}&\omega^{6}&\omega&2&\omega^{7}&\omega^{2}&\omega^{5} & 0 & 0 \\
		0&1 &2&1&2&1&2&1&2& 1 & 1 \\
		0&1 &\omega^{5}&\omega^{2}&\omega^{7}&2&\omega&\omega^{6}&\omega^{3}& 2 & 1
	\end{pmatrix}.
	$$
	Based on the Magma programe, $\mathrm{RL}_{k}\left(\mathbb{F}_{q} ,\boldsymbol{M}_{2\times 2}\right)$ is NMDS with the parameters $[11,6,5]_{3^2}$ and its weight enumerator is
	$$A(x)=1+224x^5+2352x^6+11280x^7+47000x^8+125240x^9+199824x^10+145520x^{11},$$
	which is consistent with Theorem \ref{RL2}.
\end{example}

In Theorem \ref{RL2}, we only consider the case $4\leq k\leq q-2$ and $\boldsymbol{M}_{2\times 2}\in\mathrm{GL}_{2}(\mathbb{F}_{q})$ for $p=2$. Note that, for $p=2$ and $k=3$ or $q-1$, Han and Fan \cite{A16} proved that $\mathrm{RL}_{k}\left(\mathbb{F}_{q},\boldsymbol{M}_{2}\right)$ is not NMDS. In fact, for $p=2, \boldsymbol{M}_{2\times 2}=\boldsymbol{M}_{0}$ and $k=3$ or $q-1$, we can prove that $\mathrm{RL}_{k}\left(\mathbb{F}_{q},\boldsymbol{M}_{2}\right)$ is MDS, i.e., the following Theorem \ref{k3q-1p2FqMDS}, where $\boldsymbol{M}_{0}=\left(\begin{matrix}
	0&1\\
	1&0
\end{matrix}\right)$.  

\begin{theorem}\label{k3q-1p2FqMDS}
	For $p=2$ and  $k=3$ or $q-1$, the code $\mathrm{RL}_{k}\left(\mathbb{F}_{q},\boldsymbol{M}_{0}\right)$ is MDS with the parameters $[q+2,3,q]_q$ or $[q+2,q-1,4]_q$, respectively, where $\boldsymbol{M}_{0}=\left(\begin{matrix}
		0&1\\
		1&0
	\end{matrix}\right)$.
\end{theorem}
{\bf Proof}. (1) For $p=2$ and $k=3$, by Definition \ref{EGRLdefinition}, it's easy to show that $\dim(\mathrm{RL}_{3}^{\perp}\left(\mathbb{F}_{q},\boldsymbol{M}_{0}\right))=3$ and $\mathrm{RL}_{3}\left(\mathbb{F}_{q},\boldsymbol{M}_{0}\right)$ has the generator matrix
\begin{equation}\label{RLMDS}
	\boldsymbol{G}_{3}=\begin{pmatrix}
		1 & \cdots & 1&0& 0\\
		\alpha_{1} & \cdots & \alpha_{q} & 0 & 1 \\
		\alpha_{1}^{2} & \cdots & \alpha_{q}^{2} & 1 & 0 \\
	\end{pmatrix}_{3\times (q+2)}.
\end{equation}
Furthermore, by Lemma \ref{MDSdefinition},   it's enough to prove that any $3$ columns of $\boldsymbol{G}_{3}$ are $\mathbb{F}_{q}$-linearly
independent, i.e., any  $3\times 3$ submatrices of $\boldsymbol{G}_{3}$ is nonsingular over $\mathbb{F}_{q}$. Without loss of generality, any $3\times 3$ submatrices of $\boldsymbol{G}_{3}$ has one of the following forms, 
$$\boldsymbol{G}_{31}=\begin{pmatrix}
	1&0&0\\
	\alpha_{1}& 0 & 1 \\
	\alpha_{1}^{2} & 1 & 0
\end{pmatrix}_{3\times (q+2)},\boldsymbol{G}_{32}=\begin{pmatrix}
	1&1&0\\
	\alpha_{1}&\alpha_{2}& 1 \\
	\alpha_{1}^{2} & \alpha_{2}^{2} & 0 
\end{pmatrix}_{3\times (q+2)},\boldsymbol{G}_{33}=\begin{pmatrix}
	1&1&0 \\
	\alpha_{1}&\alpha_{2}& 0 \\
	\alpha_{1}^{2} & \alpha_{2}^{2} & 1
\end{pmatrix}_{3\times (q+2)}.$$
Note that 
$$\det(\boldsymbol{G}_{31})=-1\neq 0, \det(\boldsymbol{G}_{32})=\alpha_{1}^{2}-\alpha_{2}^{2}\neq 0, \det(\boldsymbol{G}_{33})=\alpha_{2}-\alpha_{1}\neq 0,$$
which means that any $3\times 3$ submatrices of $\boldsymbol{G}_{3}$ is nonsingular over $\mathbb{F}_{q}$, i.e., $\mathrm{RL}_{3}\left(\mathbb{F}_{q},\boldsymbol{M}_{0}\right)$ is MDS. Furthermore, by the definition, $\mathrm{RL}_{3}\left(\mathbb{F}_{q},\boldsymbol{M}_{0}\right)$ is MDS with the parameters $[q+2,3,q]_{q}$. 



(2) For $p=2$ and $k=q-1$, by  Definition \ref{EGRLdefinition}, we know that $\dim\left(\mathrm{RL}_{q-1}^{\perp}\left(\mathbb{F}_{q},\boldsymbol{M}_{0}\right)\right)=q-1$ and $\mathrm{RL}_{q-1}\left(\mathbb{F}_{q},\boldsymbol{M}_{0}\right)$ has the generator matrix
\begin{equation}\label{RLMDS2}
	\boldsymbol{G}_{4}=\begin{pmatrix}
		1 & \cdots & 1&0& 0\\
		\alpha_{1} & \cdots & \alpha_{q} & 0 & 0 \\ \\
		\vdots&&\vdots&\vdots&\vdots \\
		\alpha_{1}^{q-4} & \cdots & \alpha_{q}^{q-4} & 0 & 0\\
		\alpha_{1}^{q-3} & \cdots & \alpha_{q}^{q-3} & 0 & 1\\
		\alpha_{1}^{q-2} & \cdots & \alpha_{q}^{q-2} & 1 & 0 
	\end{pmatrix}_{(q-1)\times (q+2)}.
\end{equation}
To prove that $\mathrm{RL}_{k}\left(\mathbb{F}_{q},\boldsymbol{M}_{0}\right)$ is MDS, by Lemma \ref{MDSdefinition},   it's enough to prove that any $3$ columns of $\boldsymbol{G}_{3}$ is $\mathbb{F}_{q}$-linearly
independent, i.e., any $(q-1)\times (q-1)$ submatrices of $\boldsymbol{G}_{4}$ is nonsingular over $\mathbb{F}_{q}$. Without loss of generality, any $(q-1)\times (q-1)$ submatrices of $\boldsymbol{G}_{4}$ has one of the following forms, 
$$
\boldsymbol{G}_{41}=\begin{pmatrix}   
	1 & \cdots & 1&0& 0\\
	\alpha_{1} & \cdots & \alpha_{q-3} & 0 & 0 \\
	\vdots&&\vdots&\vdots&\vdots\\
	\alpha_{1}^{q-4} & \cdots & \alpha_{q-3}^{q-4} & 0 & 0\\
	\alpha_{1}^{q-3} & \cdots & \alpha_{q-3}^{q-3} & 0 & 1\\
	\alpha_{1}^{q-2} & \cdots & \alpha_{q-3}^{q-2} & 1 & 0
\end{pmatrix},\boldsymbol{G}_{42}=\begin{pmatrix}   
1 & \cdots & 1& 0\\
\alpha_{1} & \cdots & \alpha_{q-2}  & 0 \\
\vdots&&\vdots&\vdots\\
\alpha_{1}^{q-4} & \cdots & \alpha_{q-2}^{q-4} & 0\\
\alpha_{1}^{q-3} & \cdots & \alpha_{q-2}^{q-3} & 0\\
\alpha_{1}^{q-2} & \cdots & \alpha_{q-2}^{q-2} & 1
\end{pmatrix},\boldsymbol{G}_{43}=\begin{pmatrix}   
1 & \cdots & 1& 0\\
\alpha_{1} & \cdots & \alpha_{q-2} & 0 \\
\vdots&&\vdots&\vdots\\
\alpha_{1}^{q-4} & \cdots & \alpha_{q-2}^{q-4} & 0\\
\alpha_{1}^{q-3} & \cdots & \alpha_{q-2}^{q-3} & 1\\
\alpha_{1}^{q-2} & \cdots & \alpha_{q-2}^{q-2}& 0
\end{pmatrix}.$$
Note that 
$$\det(\boldsymbol{G}_{41})=-\prod_{1\leq i<j\leq q-3}(\alpha_{j}-\alpha_{i})\neq 0,$$
$$\det(\boldsymbol{G}_{42})=\prod_{1\leq i<j\leq q-2}(\alpha_{j}-\alpha_{i})\neq 0,$$
and
$$\det(\boldsymbol{G}_{43})=(-1)^{2q-3}\sum\limits_{t=1}^{q-2}\alpha_{t}\cdot\prod_{1\leq i<j\leq q-2}(\alpha_{j}-\alpha_{i}).$$
By Lemma \ref{subsetsumneq0}, we have $\# N\left(q-2, 0,\mathbb{F}_{q}\right)=0$, it means $\sum\limits_{t=1}^{q-2}\alpha_{t}\neq 0$, and then $\det(\boldsymbol{G}_{43})\neq 0.$ Furthermore,  any $(q-1)\times (q-1)$ submatrices of $\boldsymbol{G}_{4}$ is nonsingular over $\mathbb{F}_{q}$, and so $\mathrm{RL}_{k}\left(\mathbb{F}_{q},\boldsymbol{M}_{0}\right)$ is MDS with the parameters $[q+2,q-1,4]_{q}$.

This completes the proof of Theorem $\ref{k3q-1p2FqMDS}$. 

$\hfill\Box$

\subsection{Some examples}
In this subsection, we give some examples for $\mathrm{RL}_{k}\left(\mathbb{F}_{q},\boldsymbol{M}_{2\times 2}\right)$ to be MDS or NMDS when $p=2$ and $k\in\left\{3,q-2,q-1\right\}$.

The following Example \ref{k3Fqexample1} is for Theorem \ref{k3q-1p2FqMDS}.
\begin{example}\label{k3Fqexample1}
Let $m=2,\mathbb{F}_{2^3}^{*}=\langle\omega\rangle, \alpha_{1}=0$ and $\alpha_{i}=\omega^{i-2}(i=2,\ldots,8)$, then
$$\begin{pmatrix}
	1&1 &1&1&1&1&1 & 1&0& 0\\
	0&1   &\omega&\omega^{2}&\omega^{3}&\omega^{4}&\omega^{5}&\omega^{6}&  0 & 1 \\
	0&1 &\omega^{2}&\omega^{4}&\omega^{6}&\omega&\omega^{3}&\omega^{5}&  1 & 0
\end{pmatrix}_{3\times 10}.
$$
Based on the Magma programe, $\mathrm{RL}_{k}\left(\mathbb{F}_{q} ,\boldsymbol{M}_{0}\right)$ is MDS with the parameters $[10,3,8]_{2^3}$, where $\boldsymbol{M}_{0}=\left(\begin{matrix}
	0&1\\
	1&0
\end{matrix}\right)$.
\end{example}
\begin{example}\label{RL2Fqq-1example}
	Let $m=2,\mathbb{F}_{2^3}^{*}=\langle\omega\rangle, \alpha_{1}=0$ and $\alpha_{i}=\omega^{i-2}(i=2,\ldots,8)$, then
	$$\begin{pmatrix}
		1&1 &1&1&1&1&1 & 1&0& 0\\
		0&1   &\omega&\omega^{2}&\omega^{3}&\omega^{4}&\omega^{5}&\omega^{6}&  0 & 0 \\
		0&1 &\omega^{2}&\omega^{4}&\omega^{6}&\omega&\omega^{3}&\omega^{5}&  0 & 0 \\
		0&1 &\omega^{3}&\omega^{6}&\omega^{2}&\omega^{5}&\omega&\omega^{4}&  0 & 0 \\
		0&1 &\omega^{4}&\omega&\omega^{5}&\omega^{2}&\omega^{6}&\omega^{3}&  0 & 0 \\
		0&1 &\omega^{5}&\omega^{3}&\omega&\omega^{6}&\omega^{4}&\omega^{2}& 0&1\\
		0&1 &\omega^{6}&\omega^{5}&\omega^{4}&\omega^{3}&\omega^{2}&\omega&1&0
	\end{pmatrix}_{7\times 10}.
	$$
	Based on the Magma programe, $\mathrm{RL}_{k}\left(\mathbb{F}_{q} ,\boldsymbol{M}_{0}\right)$ is MDS with the parameters $[10,7,4]_{2^3}$.
\end{example}

The following Examples \ref{k3Fqexample2}-\ref{k3Fqexample3} show that  $\mathrm{RL}_{k}\left(\mathbb{F}_{q},\boldsymbol{M}_{2\times 2}\right)$ may not be NMDS for $p=2,k=3$ and  $\boldsymbol{A}_{2\times 2}\in\mathrm{GL}_{2}(\mathbb{F}_{q})\backslash\left\{\boldsymbol{M}_{0}\right\}$, where $\boldsymbol{M}_{0}=\left(\begin{matrix}
	0&1\\
	1&0
\end{matrix}\right)$.
\begin{example}\label{k3Fqexample2}
	Let $m=2,\mathbb{F}_{2^3}^{*}=\langle\omega\rangle, \alpha_{1}=0$ and $\alpha_{i}=\omega^{i-2}(i=2,\ldots,8)$, then
	$$\begin{pmatrix}
		1&1 &1&1&1&1&1 & 1&0& 0\\
		0&1   &\omega&\omega^{2}&\omega^{3}&\omega^{4}&\omega^{5}&\omega^{6}&  0 & \omega^2 \\
		0&1 &\omega^{2}&\omega^{4}&\omega^{6}&\omega&\omega^{3}&\omega^{5}&\omega & 0 \\
	\end{pmatrix}_{3\times 10}.
	$$
	Based on the Magma programe, $\mathrm{RL}_{k}\left(\mathbb{F}_{q} ,\boldsymbol{M}_{2\times 2}\right)$ is MDS with the parameters $[10,3,8]_{2^3}$.
\end{example}

\begin{example}\label{k3Fqexample3}
Let $m=2,\mathbb{F}_{2^3}^{*}=\langle\omega\rangle, \alpha_{1}=0$ and $\alpha_{i}=\omega^{i-2}(i=2,\ldots,8)$, then
$$\begin{pmatrix} 
	1&1 &1&1&1&1&1 & 1&0& 0\\
	0&1   &\omega&\omega^{2}&\omega^{3}&\omega^{4}&\omega^{5}&\omega^{6}&  0 & \omega^2+1 \\
	0&1 &\omega^{2}&\omega^{4}&\omega^{6}&\omega&\omega^{3}&\omega^{5}&  1 & \omega
\end{pmatrix}_{3\times 10}.
$$
Based on the Magma programe, $\mathrm{RL}_{k}\left(\mathbb{F}_{q} ,\boldsymbol{M}_{2\times 2}\right)$ is NMDS with the parameters $[10,3,7]_{2^3}$.
\end{example}

The following Examples \ref{RL2Fqq-1example1}-\ref{RL2Fqq-1example2} show that  $\mathrm{RL}_{k}\left(\mathbb{F}_{q},\boldsymbol{M}_{2\times 2}\right)$ may not be NMDS for $p=2,k=q-1$ and  $\boldsymbol{M}_{2\times 2}\in\mathrm{GL}_{2}(\mathbb{F}_{q})\backslash\left\{\boldsymbol{M}_{0}\right\}$, where $\boldsymbol{M}_{0}=\left(\begin{matrix}
	0&1\\
	1&0
\end{matrix}\right)$.
\begin{example}\label{RL2Fqq-1example1}
	Let $m=2,\mathbb{F}_{2^3}^{*}=\langle\omega\rangle, \alpha_{1}=0$ and $\alpha_{i}=\omega^{i-2}(i=2,\ldots,8)$, then
	$$\begin{pmatrix}
		1&1 &1&1&1&1&1 & 1&0& 0\\
		0&1   &\omega&\omega^{2}&\omega^{3}&\omega^{4}&\omega^{5}&\omega^{6}&  0 & 0 \\
		0&1 &\omega^{2}&\omega^{4}&\omega^{6}&\omega&\omega^{3}&\omega^{5}&  0 & 0 \\
		0&1 &\omega^{3}&\omega^{6}&\omega^{2}&\omega^{5}&\omega&\omega^{4}&  0 & 0 \\
		0&1 &\omega^{4}&\omega&\omega^{5}&\omega^{2}&\omega^{6}&\omega^{3}&  0 & 0 \\
		0&1 &\omega^{5}&\omega^{3}&\omega&\omega^{6}&\omega^{4}&\omega^{2}& 0&\omega^2\\
		0&1 &\omega^{6}&\omega^{5}&\omega^{4}&\omega^{3}&\omega^{2}&\omega&\omega&0\\
	\end{pmatrix}_{7\times 10}.
	$$
	Based on the Magma programe, $\mathrm{RL}_{k}\left(\mathbb{F}_{q} ,\boldsymbol{M}_{2\times 2}\right)$ is MDS with the parameters $[10,7,4]_{2^3}$.
\end{example}
\begin{example}\label{RL2Fqq-1example2}
	Let $m=2,\mathbb{F}_{2^3}^{*}=\langle\omega\rangle, \alpha_{1}=0$ and $\alpha_{i}=\omega^{i-2}(i=2,\ldots,8)$, then
	$$\begin{pmatrix}
		
		1&1 &1&1&1&1&1 & 1&0& 0\\
		0&1   &\omega&\omega^{2}&\omega^{3}&\omega^{4}&\omega^{5}&\omega^{6}&  0 & 0 \\
		0&1 &\omega^{2}&\omega^{4}&\omega^{6}&\omega&\omega^{3}&\omega^{5}&  0 & 0 \\
		0&1 &\omega^{3}&\omega^{6}&\omega^{2}&\omega^{5}&\omega&\omega^{4}&  0 & 0 \\
		0&1 &\omega^{4}&\omega&\omega^{5}&\omega^{2}&\omega^{6}&\omega^{3}&  0 & 0 \\
		0&1 &\omega^{5}&\omega^{3}&\omega&\omega^{6}&\omega^{4}&\omega^{2}& 0&\omega^2\\
		0&1 &\omega^{6}&\omega^{5}&\omega^{4}&\omega^{3}&\omega^{2}&\omega&\omega&\omega\\
	\end{pmatrix}_{7\times 10}.
	$$
	Based on the Magma programe, $\mathrm{RL}_{k}\left(\mathbb{F}_{q} ,\boldsymbol{M}_{2\times 2}\right)$ is NMDS with the parameters $[10,7,3]_{2^3}$.
\end{example}	
\section{Conclusions}
In 2023, Han and Fan \cite{A16} proved that $\mathrm{RL}_{k}\left(\mathbb{F}_{q},\boldsymbol{M}_{0}\right)$ is NMDS when both Char($\mathbb{F}_{q}$)=2 and $4\leq k\leq q-2$ or both Char($\mathbb{F}_{q}$)$\neq 2$ and $3\leq k\leq q$. In 2025, Zhang and Zheng \cite{A13} proved that $\mathrm{RL}_{k}\left(\mathbb{F}_{q}^{*},\boldsymbol{M}_{0}\right)$ is NMDS when both Char($\mathbb{F}_{q}$)=2 and $4\leq k\leq q-3$, or both Char($\mathbb{F}_{q}$)$\neq 2$ and $3\leq k\leq q-2$. 

In this paper, by replacing $\boldsymbol{M}_{0}=\left(\begin{matrix}
	0&1\\
	1&0
\end{matrix}\right)$ with $\boldsymbol{M}_{2\times 2}=(a_{ij})\in\mathrm{GL}_{2}\left(\mathbb{F}_{q}\right)$, we improve and generalize the corresponding constructions in \cite{A13,A16} as the following table.
\begin{table}[H]
	\centering 
	\label{table_example7.1}
	\begin{tabular}{|c|c|c|c|c|c|}
		\hline
		$\boldsymbol{\alpha}$&Char($\mathbb{F}_{q}$)&$k$&$\boldsymbol{M}_{2\times 2}$&MDS&NMDS\\
		\hline
		\multirow{4}{*}{$\mathbb{F}_{q}^{*}$}&\multirow{2}{*}{even}&$3,q-2,q-1$ &$\mathrm{GL}_{2}(\mathbb{F}_{q})$&$\checkmark$(Corollary \ref{Fq0MDS}) &  \\
		\cline{3-6}
		 & &$4\leq k\leq q-3$ &$\mathrm{GL}_{2}(\mathbb{F}_{q})$&&$\checkmark$(Theorem \ref{RL1}) \\
		 \cline{2-6}
		 &\multirow{2}{*}{odd}&$q-1$ &$\mathrm{GL}_{2}(\mathbb{F}_{q})$&$\checkmark$(Corollary \ref{Fq0MDS}) &  \\
		\cline{3-6}
		& &$3\leq k\leq q-2$ &$\mathrm{GL}_{2}(\mathbb{F}_{q})$&&$\checkmark$(Theorem \ref{RL1}) \\
		\hline
		\multirow{7}{*}{$\mathbb{F}_{q}$}&\multirow{6}{*}{even}&\multirow{2}{*}{$3$} &$\boldsymbol{M}_{0}$&$\checkmark$(Theorem \ref{k3q-1p2FqMDS}) &  \\
		\cline{4-6}  
		& & &$\mathrm{GL}_{2}(\mathbb{F}_{q})\backslash\left\{\boldsymbol{M}_{0}\right\}$& Example \ref{k3Fqexample2}&Example \ref{k3Fqexample3}\\
		\cline{3-6}
		& & \multirow{2}{*}{$4\leq k\leq q-2$} &$\boldsymbol{M}_{0}$&$\checkmark$( Theorem \ref{RL2})& \\
		\cline{4-6}
		& & &$\mathrm{GL}_{2}(\mathbb{F}_{q})\backslash\left\{\boldsymbol{M}_{0}\right\}$&$\checkmark$(Theorem \ref{RL2}) &\\
		\cline{3-6}
		& & \multirow{2}{*}{$q-1$} &$\boldsymbol{M}_{0}$&$\checkmark$(Theorem \ref{k3q-1p2FqMDS})&  \\
		\cline{4-6}
		& &	&$\mathrm{GL}_{2}(\mathbb{F}_{q})\backslash\left\{\boldsymbol{M}_{0}\right\}$&Example \ref{RL2Fqq-1example1}& Example \ref{RL2Fqq-1example2}\\
		\cline{2-6}
		& odd&$3\leq k\leq q$ &$ \mathrm{GL}_{2}(\mathbb{F}_{q})$&$\checkmark$(Theorem \ref{RL2}) &  \\
		\hline 
	\end{tabular}
\end{table}






\begin{thebibliography}{1}
\bibitem{A1} Kløve T. Codes for Error Detection (Series on Coding Theory and Cryptology; V. 2)[M]. World Scientific, 2007.
\bibitem{A2} S. Doudunekov, I. Landjev, On near MDS codes, J. Geom. 54 (1–2) (1994) 30–43.
\bibitem{A3} W. Huffman, V. Pless, Fundamentals of Error Correcting Codes, Cambridge University Press, Cambridge, 2003.
\bibitem{A4} F. MacWilliams, N. Sloane, The Theory of Error-Correcting Codes, North Holland, Amsterdam, 1977.
\bibitem{A5} D. Simos, Z. Varbanov, MDS codes, NMDS codes and their secret-sharing schemes, Accessed April 2018.
\bibitem{A6} A. Thomas, B. Rajan, Binary informed source codes and index codes using certain near MDS codes, IEEE Trans. Commun. 66 (5) (2018) 2181–2190.
\bibitem{A7} Li Y, Zhu S. New non-GRS type MDS codes and NMDS codes[J]. arXiv preprint arXiv:2401.04360, 2024.
\bibitem{A8} Wu Y, Heng Z, Li C, et al. More MDS codes of non-Reed-Solomon type[J]. arXiv preprint arXiv:2401.03391, 2024.
\bibitem{A9} Jin L, Ma L, Xing C, et al. New families of non-Reed-Solomon MDS codes[J]. arXiv preprint arXiv:2411.14779, 2024.
\bibitem{A10} Xu L, Fan C, Han D. Near-MDS codes from maximal arcs in PG (2, q)[J]. Finite Fields and Their Applications, 2024, 93: 102338.
\bibitem{A11} Fan C, Wang A, Xu L. New classes of NMDS codes with dimension 3[J]. Designs, Codes and Cryptography, 2024, 92(2): 397-418.
\bibitem{A12} Ding Y, Li Y, Zhu S. Four new families of NMDS codes with dimension 4 and their applications[J]. Finite Fields and Their Applications, 2024, 99: 102495.
\bibitem{A13} Zhang Y, Zheng D, Wang X, et al. Several new infinite families of NMDS codes with arbitrary dimensions supporting $ t $-designs[J]. arXiv preprint arXiv:2504.06546, 2025.
\bibitem{A14} Han D, Zhang H. Explicit constructions of NMDS self-dual codes[J]. Designs, Codes and Cryptography, 2024, 92(11): 3573-3585.
\bibitem{A15} Yin Y, Yan H. Constructions of several families of MDS codes and NMDS codes[J]. Advances in Mathematics of Communications, 2024: 0-0.
\bibitem{A16} Han D, Fan C. Roth–Lempel NMDS codes of non-elliptic-curve type[J]. IEEE Transactions on Information Theory, 2023, 69(9): 5670-5675.
\bibitem{A17} Heng Z, Li X. Near MDS codes with dimension 4 and their application in locally recoverable codes[C]//International Workshop on the Arithmetic of Finite Fields. Cham: Springer International Publishing, 2022: 142-158.
\bibitem{A23} Abdukhalikov K, Ding C, Verma G K. Some constructions of non-generalized Reed-Solomon MDS Codes[J]. arXiv preprint arXiv:2506.04080, 2025.
\bibitem{A18} R. M. Roth and A. Lempel, “A construction of non-Reed–Solomon type
MDS codes,” IEEE Trans. Inf. Theory, vol. 35, no. 3, pp. 655–657,
May 1989.
\bibitem{A19} Xu L, Fan C. Near MDS codes of non-elliptic-curve type from Reed-Solomon codes[J]. Discrete Mathematics, 2023, 346(9): 113490.
\bibitem{A20} Li J, Wan D. On the subset sum problem over finite fields[J]. Finite Fields and Their Applications, 2008, 14(4): 911-929. 
\bibitem{A21} Zhu C, Liao Q. The (+)-extended twisted generalized Reed-Solomon code[J]. Discrete Mathematics, 2024, 347(2): 113749.
\bibitem{A22} Wu Y, Heng Z, Li C, et al. More MDS codes of non-Reed-Solomon type[J]. arXiv preprint arXiv:2401.03391, 2024.

\end{thebibliography}
\end{document}